\documentclass[reprint,showpacs,amsmath,amssymb,aps]{revtex4-1}

\usepackage{graphicx}
\usepackage{dcolumn}
\usepackage{bm}

\usepackage{aas_macros}

\begin{document}

\title{Constraining Galactic dark matter with gamma-ray pixel counts statistics}

\author{H.-S. Zechlin}
\email{zechlin@to.infn.it}
\affiliation{Istituto Nazionale di Fisica Nucleare, Sezione di Torino, via P. Giuria, 1, I-10125 Torino, Italy}
\author{S. Manconi}
\email{manconi@to.infn.it}
\author{F. Donato}
\email{donato@to.infn.it}
\affiliation{Dipartimento di Fisica, Universit\`a di Torino, via P. Giuria, 1, I-10125 Torino, Italy}
\affiliation{Istituto Nazionale di Fisica Nucleare, Sezione di Torino, via P. Giuria, 1, I-10125 Torino, Italy}

\begin{abstract}
Gamma-ray searches for new physics such as dark matter are often
driven by investigating the composition of the extragalactic gamma-ray
background (EGB). Classic approaches to EGB decomposition manifest in
resolving individual point sources and dissecting the intensity
spectrum of the remaining unresolved component. Furthermore,
statistical methods have recently been proven to outperform the
sensitivity of classic source detection algorithms in finding
point-source populations in the unresolved flux regime. In this
article, we employ the 1-point photon count statistics of eight years
of \emph{Fermi}-LAT data to resolve the population of extragalactic
point sources and to decompose the diffuse isotropic background
contribution for Galactic latitudes $|b|\geq30^\circ$. We use three
adjacent energy bins between 1 and 10\,GeV.  For the first time, we
extend the analysis to incorporate a potential contribution from
annihilating dark matter smoothly distributed in the Galaxy.  We
investigate the sensitivity reach of 1-point statistics for
constraining the thermally-averaged self-annihilation cross section
$\langle \sigma v \rangle$ of dark matter, using different template
models for the Galactic foreground emission.  Given the official {\it
  Fermi}-LAT interstellar emission model, we set upper bounds on the
DM self-annihilation cross section $\langle \sigma v \rangle$ that are
comparable with the constraints obtained by other indirect detection
methods, in particular by the stacking analysis of several dwarf
spheroidal galaxies.
\end{abstract}


\maketitle


\section{\label{sec:intro}Introduction}
The Large Area Telescope (LAT;
\cite{2009ApJ...697.1071A,2012ApJS..203....4A}) on board the
\emph{Fermi} Gamma-ray Space Telescope (\emph{Fermi}) has led to
tremendous progress in understanding the nature of non-thermal gamma
rays reaching the Earth. In general, the all-sky gamma-ray emission is
composed of two contributions: a bright foreground produced in our own
Galaxy, and all other gamma rays originating from farther sources
outside the Galaxy. The latter contributions mutually accumulate to
the extragalactic gamma-ray background (EGB, see
\cite{2015PhR...598....1F}). The EGB itself disseminates into point
sources (PSs) and an effectively isotropic diffuse background
contribution, the IGRB. At Galactic latitudes $|b|>20^\circ$, the EGB
has been detected with the LAT between 100\,MeV and 820\,GeV with
unprecedented precision \cite{2015ApJ...799...86A}.

Blazars represent the brightest and most numerous source population
among all sources resolved in the EGB (e.g.,
\cite{2009ApJ...702..523I,2011ApJ...743..171A,2011PhRvD..84j3007A,
  2012ApJ...751..108A,2012ApJ...753...45S}). In addition, gamma-ray
source catalogs such as the \emph{Fermi} Large Area Telescope Third
Source Catalog (3FGL; \cite{2015ApJS..218...23A}) list associations of
gamma-ray sources with other source types, among them misaligned
active galactic nuclei and star-forming galaxies (see
\cite{Massaro2015} and references therein).  Different source
populations distinguish themselves observationally by spectral and
temporal characteristics. Another population intrinsic quantity is
their source count distribution $\mathrm{d}N/\mathrm{d}S$, denoting
the number of sources $N$ per solid angle element $\mathrm{d}\Omega$
with integral fluxes in the interval $(S,S+\mathrm{d}S)$. The
phenomenology of $\mathrm{d}N/\mathrm{d}S$ distributions can be
obtained from specific, mostly data-driven models, derived from basic
principles of source-intrinsic gamma-ray production and cosmological
source evolution. The extrapolation of these models suggests the
existence of numerous sources with fluxes fainter than current source
detection thresholds
\cite{2011ApJ...733...66I,2012ApJ...753...45S,2014ApJ...786..129D,
  2014ApJ...780..161D, 2014ApJ...796...14C, 2014JCAP...09..043T,
  2015ApJ...800L..27A}. Thus, the IGRB is expected to originate at
least partly from unresolved faint point-source populations. As
opposed to source contributions, the IGRB might contain purely diffuse
components, among them diffuse gamma rays originating from cosmic-ray
interactions with the intergalactic medium \cite{2015ApJ...799...86A}
or the annihilation or decay of dark matter (DM) particles in the
Milky~Way's halo and halos of outer galaxies. To that regard, the IGRB
serves as a complementary probe for Galactic and cosmological DM
\cite{Ullio:2002pj, Abazajian:2010zb, Papucci:2009gd,
  2010NuPhB.840..284C, Calore:2013yia,DiMauro:2015tfa}.

Singling out a possible DM contribution to the IGRB usually is
complicated by a number of uncertainties: The modeling of Galactic
diffuse foreground emission, the uncertain contribution from
unresolved sources, and the accuracy of model predictions for the DM
signal strength itself prevent the IGRB from currently being
considered a clean target for DM searches
\cite{DiMauro:2015tfa,Ackermann:2015tah}. However, the sensitivity of
DM searches in the IGRB with \textit{Fermi}-LAT data has been
demonstrated to reach competitive limits, for example with respect to
recent dwarf spheroidal galaxy observations
\cite{Calore:2013yia,DiMauro:2015tfa,Ackermann:2015tah}. In the near
future, both improvements of the LAT sensitivity for detecting
point-like sources and various attempts of reducing systematic model
uncertainties with complementary observations will consolidate the
crucial role of the IGRB in DM searches.

The dissection of the EGB by means of individual source detections and
intensity measurements can be complemented by statistical
methods. Analyses employing statistical properties of the observed
counts map have been demonstrated to be capable of measuring
$\mathrm{d}N/\mathrm{d}S$ and the diffuse EGB components
\cite{2009PhRvD..80h3504D,2009JCAP...07..007L,2011ApJ...738..181M,
  2015JCAP...09..027F,2015PhRvD..91h3539M,
  2015A&A...581A.126S,2016PhRvL.116e1103L,2016ApJ...832..117L,
  2016ApJS..225...18Z}. We have shown in
Refs.~\cite{2016ApJS..225...18Z} and \cite{2016ApJ...826L..31Z}
(henceforth Z16a and Z16b, respectively) that the 1-point probability
distribution function (1pPDF) of counts maps serves as a unique tool
for precise measurements of $\mathrm{d}N/\mathrm{d}S$ and the EGB's
composition. In short, the 1pPDF represents the probability
distribution function of photon counts as distributed in a pixelized
sky map. Statistical measurements are not only complementary to
standard analysis procedures, but they also significantly increase the
sensitivity for resolving faint point-source populations.

Searches for DM signals in the EGB with statistical methods such as
the 1pPDF will thus particularly profit from a better sensitivity with
regard to resolving faint point sources as well as from the unique
dissection capabilities in general. In this article, we extend the
1pPDF method presented in Z16a,b to incorporate an additional
component representing a smooth Galactic DM halo. We further
investigate the achievable sensitivity reach of the 1pPDF method with
regard to constraining the DM self-annihilation cross section $\langle
\sigma v \rangle$, on the basis of eight years of \emph{Fermi}-LAT
data between 1 and 10\,GeV and different Galactic foreground emission
models. Our exploration is meant as a sensitivity study to seek
diffuse photons originating from DM annihilation at high Galactic
latitudes. Since the 1pPDF method incorporates both resolved and
unresolved point sources, source locations do not have to be excluded
from the data and results can be considered as catalog independent.

The 1pPDF method and the modeling of the gamma-ray sky (including the
DM component) are discussed in
Section~\ref{sec:1pPDF}. \emph{Fermi}-LAT data reduction and the data
analysis setup are discussed in
Section~\ref{sec:data}. Section~\ref{sec:data_analysis} focuses on the
actual 1pPDF analysis, while results are discussed in
Section~\ref{sec:results}. The paper concludes with
Section~\ref{sec:conclusions}.

\section{\label{sec:1pPDF}The 1pPDF Method and DM}
The mathematical foundations of the 1pPDF method, the implementation,
and the application of the method to \emph{Fermi}-LAT data are
discussed in Z16a,b. In this article, we extend the model of the
gamma-ray sky to an additional component, i.e. the total gamma-ray
emission is described by superimposing four different contributions:
\begin{enumerate}
\item An isotropic distribution of gamma-ray point sources, described
  with a differential source-count distribution
  $\mathrm{d}N/\mathrm{d}S$. The $\mathrm{d}N/\mathrm{d}S$
  distribution is parameterized with a multiply broken power law
  (MBPL), with its free fit parameters comprising the overall
  normalization, a number of $N_\mathrm{b}$ break positions, and
  therefore $N_\mathrm{b}+1$ power-law components connecting the
  breaks.
\item A diffuse component of Galactic foreground emission, described
  with an interstellar emission model (IEM). Further details on the
  considered IEMs are given in Section~\ref{ssec:IEMs}. The global
  normalization of the IEM template is kept a free fit parameter,
  $A_\mathrm{gal}$.
\item A diffuse component accounting for all contributions
  indistinguishable from purely diffuse isotropic emission. The
  diffuse isotropic background emission is assumed to follow a power
  law spectrum (photon index $\Gamma=2.3$), with its integral flux
  $F_\mathrm{iso}$ serving as the free normalization parameter.
\item A distribution of Galactic DM, representing a typical smooth DM
  halo. The gamma-ray emission from the DM halo is included as a
  template with a free global normalization parameter,
  $A_\mathrm{DM}$. Details are discussed in
  Section~\ref{ssec:dm_model}.
\end{enumerate}
In the subsequent paragraphs, the mathematical base of the 1pPDF is
briefly revisited. The reader is referred to Z16a, Section~2, for
details.

Let $\mathcal{P}^{(p)} (t)$ be a probability generating function of a
discrete probability distribution $p_k^{(p)}$, where $t\in\mathbb{R}$
is an auxiliary variable, $k = 0,1,2,\dots$ is a discrete random
variable, and $p$ denotes the evaluated map pixel $p$. Then
$p_k^{(p)}$ is given by
\begin{equation}\label{eq:pkcalc}
p^{(p)}_k = \frac{1}{k!} \left. \frac{\mathrm{d}^k \mathcal{P}^{(p)} (t)}
{\mathrm{d}t^k}\right|_{t=0} .
\end{equation}
The generic representation of the generating function for photon count
maps can be derived from a superposition of Poisson processes:
\begin{equation}\label{eq:gfgen1}
\mathcal{P}^{(p)} (t) = \exp \left[ \sum_{m=1}^{\infty} x^{(p)}_m 
\left( t^m -1 \right) \right] ,
\end{equation}
where $x^{(p)}_m$ is the expected number of point sources per pixel
$p$ that contribute exactly $m$ photons to the total pixel photon
content. The quantity $x^{(p)}_m$ is therefore given by the
source-count distribution $\mathrm{d}N/\mathrm{d}S$, where $S$ denotes
the integral photon flux of a source in the energy band
$[E_\mathrm{min},E_\mathrm{max}]$, i.e.
\begin{equation}\label{eq:xm}
x^{(p)}_m = \Omega_\mathrm{pix} \int_0^\infty \mathrm{d}S 
\frac{\mathrm{d}N}{\mathrm{d}S} \frac{(\mathcal{C}^{(p)}\!(S) )^m}{m!} 
e^{-\mathcal{C}^{(p)}\!(S)} ,
\end{equation}
where $\mathcal{C}^{(p)}\!(S)$ denotes the average number of photons
contributed to pixel $p$ by a source with flux $S$, and
$\Omega_\mathrm{pix}$ is the solid angle of the pixel.  Diffuse
background components can be represented by 1-photon source terms,
i.e.
\begin{equation}\label{eq:Dgen}
\mathcal{D}^{(p)} (t) = \exp \left[ x_\mathrm{diff}^{(p)}\,(t-1) \right] ,
\end{equation}
where $x^{(p)}_\mathrm{diff}$ denotes the number of diffuse photon
counts expected in map pixel $p$.  The total generating function then
factorizes in the point-source component and the diffuse component,
\begin{equation}\label{eq:Ptot}
\mathcal{P}^{(p)} (t) = \mathcal{P}_\mathrm{S}^{(p)} (t)\,\mathcal{D}^{(p)} (t)\,.
\end{equation}

For our model of the gamma-ray sky, the total diffuse contribution
$x_\mathrm{diff}^{(p)}$ is given by
\begin{equation}\label{eq:xdiff}
  x_\mathrm{diff}^{(p)} = A_\mathrm{gal} x_\mathrm{gal}^{(p)} +
  A_\mathrm{DM} x_\mathrm{DM}^{(p)} + \frac{x_\mathrm{iso}^{(p)}}{F_\mathrm{iso}} F_\mathrm{iso}\,,
\end{equation}
where $x_\mathrm{b}^{(p)}$, with $\mathrm{b} \in
\{\mathrm{gal,DM,iso}\}$, reads
\begin{equation}\label{eq:xb}
x^{(p)}_\mathrm{b} = \int_{\Omega_\mathrm{pix}} \,\mathrm{d}\Omega
\int_{E_\mathrm{min}}^{E_\mathrm{max}} 
\mathrm{d}E\,f^{(p)}_\mathrm{b}(E)\,\mathcal{E}^{(p)} (E)\, ,
\end{equation}
with $f^{(p)}_\mathrm{b}(E)$ being the differential flux as function
of the energy $E$, and $\mathcal{E}^{(p)} (E)$ the pixel-dependent
exposure of the detector.

The likelihood function is given by a product over the probabilities
$P$ of finding the number $k_p$ of measured counts in pixel $p$. With
the probability $P(k_p)$ directly given by Eq.~\ref{eq:pkcalc}, the
total likelihood of the region of interest (ROI), covered by
$N_\mathrm{pix}$ pixels, reads
\begin{equation}\label{eq:likelihood}
\mathcal{L} ({\bf \Theta}) = \prod_{p=1}^{N_\mathrm{pix}} P(k_p)
\end{equation}
for a given parameter vector ${\bf \Theta}$.

For real data sets, we note that Eq.~\ref{eq:xm} has to be corrected
for source-smearing effects caused by a finite detector point-spread
function (PSF). See Z16a for details.

\subsection{\label{ssec:IEMs}Galactic Foreground Emission}
Gamma rays from the interaction of cosmic rays (CRs) with interstellar
gas and interstellar radiation fields (IRFs) in our Galaxy are the
main contributors to the emission observed with the LAT above
$100$~MeV and constitute the diffuse Galactic foreground. This
emission provides a complementary tool to study the properties of CRs
throughout the Galaxy and the interstellar medium, and dominates the
emission coming from point sources, extragalactic diffuse
contributions, as well as a possible DM signal.

Modeling the diffuse Galactic emission is complex and equipped with
high systematic uncertainties. The morphological structure and the
spectrum of the Galactic emission are caused by a sum of different
contributions, driven by a variety of different physical
parameters. The interaction of CRs with the interstellar HI and $H_2$
gas is responsible for the production of gamma rays through
non-thermal bremsstrahlung and $\pi^0$ production and their subsequent
decay, while the interaction of Galactic electrons with the IRFs
produces gamma rays through inverse Compton (IC) scattering. To
properly compute these contributions, models for CR sources, injection
spectra, and diffusion in our Galaxy, as well as a good understanding
of the interstellar gas distribution and the structure of radiation
fields are required. The computation of the different components is
additionally hampered, for instance, by the presence of large-scale
structures correlated to Galactic Loop~I \cite{Wolleben:2007pq} or the
\emph{Fermi} Bubbles \cite{Fermi-LAT:2014sfa}, or by well-known
degeneracies among propagation parameters.

A search for DM (and other additional components) with methods relying
on IEM templates requires a rigorous assessment of possible
systematics driven by the IEM.  In particular the IC component, which
is expected to be a possible cause of degeneracies with the DM
component, would need to be freely adjustable by the fit,
cf. Ref.~\cite{2012ApJ...761...91A}. Nevertheless, when considering
high latitude regions ($|b| > 30^\circ$) away from the Galactic plane
(GP), a representative small set of IEMs may be considered sufficient
to quantify uncertainties due to Galactic foreground modeling
\cite{2015ApJ...799...86A,Charles:2016pgz}.  To bracket the
uncertainties inherent to the IEM we consider 4 different models: Our
benchmark IEM adopts the official spatial and spectral template as
provided by the \emph{Fermi}-LAT Collaboration for the \texttt{Pass 8}
analysis framework ({\tt gll\_iem\_v06.fits}, see
Ref.~\cite{2016ApJS..223...26A} and Section~\ref{sec:data}). In
addition, we compare analyses using the models A, B, and C as used in
Ref.~\cite{2015ApJ...799...86A} to bracket the systematic
uncertainties of the IGRB analysis. The same three models A, B, C have
also been used in Refs.~\cite{Ackermann:2015tah,DiMauro:2015tfa} to
study uncertainties related to the diffuse Galactic emission modeling
when searching for DM contributions in the IGRB data. An extended
description of the characteristics of the models A, B, C can be found
in Appendix A of Ref.~\cite{2015ApJ...799...86A}. Here, we summarize
the main elements and differences between the models, in particular
focusing on Galactic latitudes $|b| > 30^\circ$.

\begin{figure}
\centering
\includegraphics[width=\columnwidth]{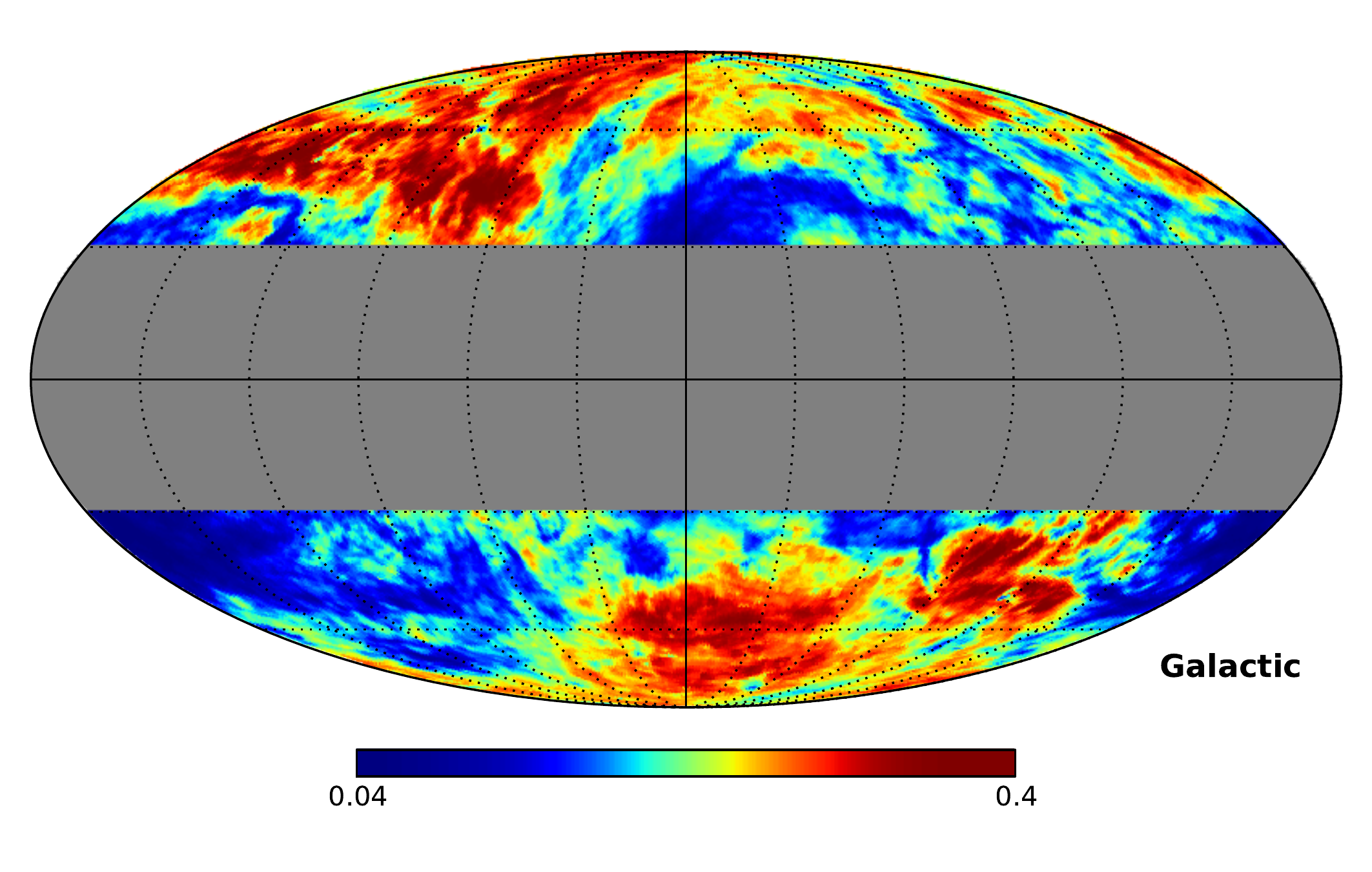}
\caption{\label{fig:modAB} Relative difference between the Galactic
  diffuse emission predicted by model A and B, for latitudes $|b| >
  30^\circ$. Models are taken from Ref.~\cite{2015ApJ...799...86A} and
  are integrated in the energy bin between 1.99 and 5.0\,GeV. The
  Mollweide projection is given in Galactic coordinates, centered on
  the GC. The GP region has been masked in gray.}
\end{figure}

The A, B, C model templates provided for the emission related to HI
+$H_2$ and IC have been obtained with a modified version of the
GALPROP propagation code (see Ref.~\cite{2015ApJ...799...86A} for
details). The main differences between these models regard the IC
component and the treatment of CR diffusion. In model B, an additional
electron source population near the Galactic Center (GC) produces the
bulk of the IC emission. With respect to model A and C, this
translates into a better agreement between the template prediction for
the IC component and the fit of model B to gamma-ray data for Galactic
latitudes $|b| > 20^\circ$, see Ref.~\cite{2015ApJ...799...86A} for
details. In fact, we found that the relative difference between the IC
component in the energy bin between 1.99 and 5.0\,GeV and $|b| >
30^\circ$ for models A and B is (IC$_{A}$ $-$ IC$_{B}$)/IC$_{A} > 0$
for all pixels, ranging from 40\% to 70\%. The same relative
difference between the IC components of models B and C ranges from
60\% up to a factor of 2 in the outer Galaxy. Fig.~\ref{fig:modAB}
represents the relative difference between the entire emission
predicted by model A and B, in the energy bin between 1.99 and
5.0\,GeV for latitudes $|b| > 30^\circ$. The differences follow the
complicated structure of Galactic gas and indicate that model A
predicts higher (40\% at most) diffuse emission in the whole ROI. In
model A and B, the CR diffusion coefficient and re-acceleration
strength are constant throughout the Galaxy. In model C, instead, a
dependency on the galactocentric radius and height is introduced,
causing a more efficient transport of CRs and therefore higher
gamma-ray intensities in the outer Galaxy, as shown by Fig.~20 in
Ref.~\cite{2015ApJ...799...86A}.

In our analysis, each IEM is re-normalized with an additional global
normalization factor $A_\mathrm{gal}$ that is allowed to float
freely. We underline that the various models are studied here to
explore the effect of changing foreground morphology, in particular of
the IC emission, on the contribution from the additional Galactic DM
component. A complete study on whether the data prefer one of those
models over the other is beyond the scope of this work, and it is
extensively addressed for example in Ref.~\cite{2012ApJ...750....3A}.

\subsection{\label{ssec:dm_model}The DM Component}
Based upon the assumption that the building blocks of DM are new
fundamental particles, e.g., weakly interacting massive particles
(WIMPs), DM can self-annihilate or decay into standard model final
states. Gamma-ray photons are then unavoidably produced by secondary
processes such as hadronization, the subsequent decay of
$\pi^0$-mesons, and internal bremsstrahlung, which lead to a
continuous gamma-ray spectrum over several decades in energy, as well
as direct annihilation into line-like features.  The observed
differential gamma-ray flux per unit energy interval
$(E,E+\mathrm{d}E)$ and solid angle $\mathrm{d}\Omega$ from DM
annihilation in a given celestial direction reads
\begin{equation}\label{eq:dmflux}
  \frac{\mathrm{d}\phi_\mathrm{DM}}{\mathrm{d}E \mathrm{d}\Omega} =
  \frac{1}{4\pi} \frac{\langle \sigma v \rangle}{2}
  r_\odot \frac{\rho_\odot^2}{m^2_\mathrm{DM}} \sum_f
  \left(\frac{\mathrm{d}N_f}{\mathrm{d}E} B_f \right) \mathcal{J}(\psi)\,.
\end{equation}
The quantity $\langle \sigma v \rangle$ resembles the
thermally-averaged self-annihilation cross section times the relative
velocity, $m_\mathrm{DM}$ denotes the DM particle mass, and $r_\odot =
8.5\,\mathrm{kpc}$ and $\rho_\odot = 0.4$\,GeV\,cm$^{-3}$
\cite{Catena:2009mf,Read:2014qva} are normalization constants,
i.e. the galactocentric Solar distance and the DM density at
$r_\odot$, respectively. Equation~\ref{eq:dmflux} is valid for
self-conjugated DM particles. The differential gamma-ray spectrum
yielded by DM annihilation into the standard model final state $f$
with branching ratio $B_f$ is given by
$\mathrm{d}N_f/\mathrm{d}E$. The dimensionless J-factor reads
\begin{equation}\label{eq:j_factor}
  \mathcal{J}(\psi) = \frac{1}{r_\odot} \int_\mathrm{los}
  \left( \frac{\rho [r(l)]}{\rho_\odot} \right)^2 \mathrm{d}l(\psi)\,.
\end{equation}
Here, $\rho(r)$ denotes the DM density profile as a function of the
galactocentric radius $r$, and the line-of-sight (los), $l$, as
measured from the Galactic position of the Sun is given by $r(l,
\psi)= \sqrt{r_\odot^2 + l^2 -2r_\odot l \cos\psi}$, where $\psi$ is
the angle between the vector pointing to the GC and the direction of
observation.

\begin{figure}
\centering
\includegraphics[width=\columnwidth]{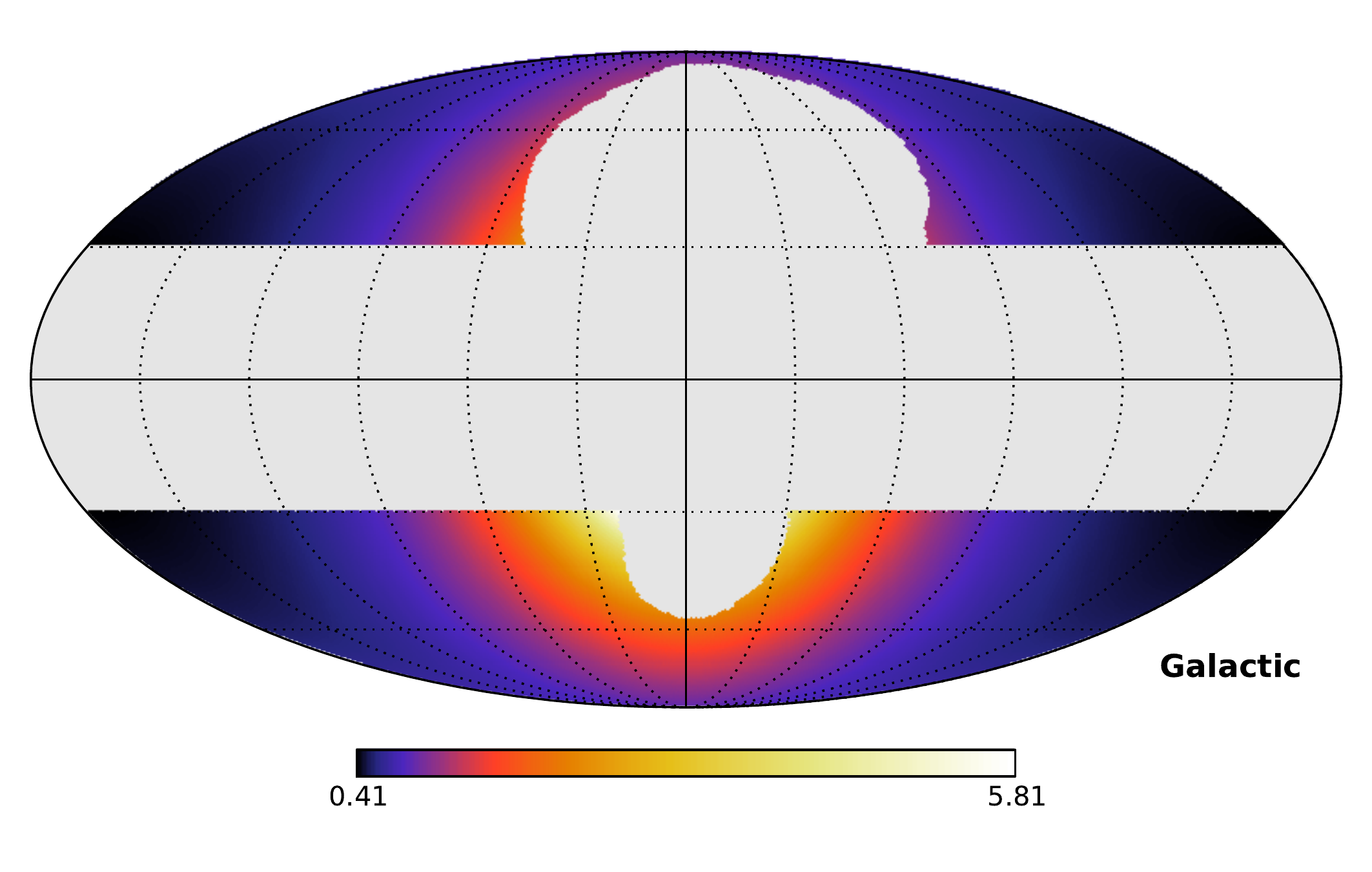}
\caption{\label{fig:j_factor}Dimensionless J-factor for annihilating
  DM as distributed in the Galaxy following an Einasto profile.  The
  Mollweide projection is given in Galactic coordinates, centered on
  the position of the GC. The GP and the regions covered by the Fermi
  Bubbles and Galactic Loop~I have been masked in gray,
  cf. Section~\ref{sec:data_analysis}.}
\end{figure}

We consider here the contribution from DM annihilation in a smooth
Galactic halo. We neglect a possible contribution from Galactic DM
subhalos, which can be modeled as point-like or slightly extended
sources in almost all relevant DM scenarios \citep[see,
  e.g.,][]{2012A&A...538A..93Z,2012ApJ...747..121A,2012JCAP...11..050Z,
  2015JCAP...12..035B,2016JCAP...05..028S,2017JCAP...04..018H,Calore:2016ogv}
and will therefore contribute to the generic $\mathrm{d}N/\mathrm{d}S$
component. For the density profile of the smooth Galactic halo, we
consider an Einasto profile \cite{1965TrAlm...5...87E}
\begin{equation}\label{eq:ein}
  \rho(r)= \rho_\odot \exp \left( -\frac{2}{\alpha}
  \frac{r^\alpha - r_\odot^\alpha}{r_s^\alpha}\right)\,,
\end{equation}
with $\alpha = 0.17$ and $r_s = 21.8$\,kpc
\cite{Bringmann:2012ez}. The dimensionless J-factor template map was
generated by solving Eq.~\ref{eq:j_factor} independently for each map
pixel $p$, see Fig.~\ref{fig:j_factor}. The HEALPix resolution of the
template map was chosen corresponding to the resolution used in the
data analysis, see Section~\ref{sec:data}.

The quantity $x^{(p)}_\mathrm{DM}$ (see Eq.~\ref{eq:xdiff}) is given
by Eqs.~\ref{eq:xb} and \ref{eq:dmflux}. Here, we assume a benchmark
annihilation cross section of $\langle \sigma v \rangle_0 =
10^{-26}\,\mathrm{cm}^3\mathrm{s}^{-1}$, such that the dimensionless
fit parameter $A_\mathrm{DM}$ represents a rescaling of the DM flux
provided by Eq.~\ref{eq:dmflux}, given $\langle \sigma v \rangle_0$
and the chosen normalization $\rho_\odot$. We consider DM annihilation
into pure $b\overline{b}$ quark final states and pure $\tau^+\tau^-$
lepton final states. Those channels serve as benchmark annihilation
channels, bracketing general DM annihilation scenarios. The gamma-ray
spectra emerging from the final states were taken from
Ref.~\cite{2011JCAP...03..051C}. Possible secondary IC emission from
the scattering of charged light leptons with IRFs was neglected
\cite{2010NuPhB.840..284C}. The peak of the energy spectra $E^2
\mathrm{d}{N}/\mathrm{d}E$ usually scales with the DM particle mass,
$E_\mathrm{peak} \propto m_\mathrm{DM}$, implying that the best choice
for the energy bin to analyze can depend on the DM mass. We
investigated DM particles with masses between 5\,GeV and 1\,TeV.

\section{\label{sec:data}Fermi-LAT Data Reduction}
We processed all-sky \emph{Fermi}-LAT gamma-ray data that were taken
within the first 8 years of the mission, i.e. from 2008 August 4
(239,557,417\,s MET) through 2016 August 4 (492,018,220\,s MET). We
used \texttt{Pass 8} data \footnote{\emph{Fermi}-LAT data are publicly
  available at
  \texttt{https://heasarc.gsfc.nasa.gov/FTP/fermi/data/lat/\\weekly/photon/}}
along with the corresponding instrument response functions. The Fermi
Science Tools (v10r0p5, released date 2015 June
24) \footnote{\texttt{https://fermi.gsfc.nasa.gov/ssc/data/analysis/\\software/}}
were employed for event selection and data processing.

To reduce systematic uncertainties, the 1pPDF analysis requires clean
data sets with low residual cosmic-ray backgrounds and event samples
exhibiting comparably mild PSF smoothing effects (see Z16a,b). We
therefore only used events passing the most stringent \texttt{Pass 8}
data classification criteria, i.e. belonging to the
\texttt{ULTRACLEANVETO} event class. The corresponding instrument
response functions \texttt{P8R2\_ULTRACLEANVETO\_V6} were
used. Furthermore, we restricted the event sample to the \texttt{PSF3}
quartile, to avoid significant PSF smoothing. A possible contamination
from the Earth's limb was reduced by restricting the zenith angle to a
maximum of $90^\circ$. The data selection referred to standard quality
selection criteria (\texttt{DATA\_QUAL==1} and
\texttt{LAT\_CONFIG==1}), and the rocking angle of the satellite was
constrained to values smaller than $52^\circ$.

To maximize the sensitivity for the $m_\mathrm{DM}$ parameter space
(see Section~\ref{ssec:dm_model}), we chose to analyze three adjacent
energy bands: (i) 1.04--1.99\,GeV, (ii) 1.99--5.0\,GeV, and (iii)
5.0--10.4\,GeV. The bands were selected following Z16b. The PSF as a
function of energy becomes significantly larger than $1^\circ$ at
energies below $\sim1$\,GeV, but approaches sizes below $0.1^\circ$
for energies above $\sim10$\,GeV. The effective PSF of each energy
band was derived by weighting the PSF with the average exposure
$\mathcal{E}(E)$ of the ROI and power-law-type energy spectra (see
Eq.~19 in Z16a). The effective PSF widths corresponding to the three
energy bands are (i) $0.31^\circ$, (ii) $0.18^\circ$, and (iii)
$0.10^\circ$.

The data were spatially binned using the HEALPix equal-area
pixelization scheme \cite{2005ApJ...622..759G}. Thus, the entire sky
is covered by $N_\mathrm{pix}=12 N^2_\mathrm{side}$ pixels, where
$N_\mathrm{side} = 2^\kappa$. We compared two choices for the
resolution parameter $\kappa$ of the pixelization, i.e. $\kappa=6$ and
$\kappa=7$, approximating the size of the PSF.

\section{\label{sec:data_analysis}Data Analysis}
The 1pPDF likelihood function $\mathcal{L}({\bf \Theta})$ as defined
in Eq.~\ref{eq:likelihood} was analyzed following the method of
Z16a. We used the Bayesian Markov Chain Monte Carlo (MCMC) sampler
\texttt{MultiNest} \cite{2008MNRAS.384..449F,2009MNRAS.398.1601F} to
sample the posterior distribution $P({\bf \Theta}) = \mathcal{L}({\bf
  \Theta}) \pi({\bf \Theta}) / \mathcal{Z}$, where $\pi({\bf \Theta})$
is the prior and $\mathcal{Z}$ is the Bayesian
evidence. \texttt{MultiNest} was operated in its standard
configuration. We used 1000 live points together with a tolerance
criterion of 0.2. The configuration was checked for stability. Priors
were either log-flat or flat, cf. Z16a, and their ranges were chosen
such that to sufficiently cover the posterior distributions. In
particular, the newly introduced prior for $A_\mathrm{DM}$ was of the
log-flat type.

From the final posterior sample, we built one-dimensional profile
likelihood functions \cite{2005NIMPA.551..493R} for each parameter, in
order to get prior-independent frequentist parameter
estimates. Best-fit parameter estimates refer to the maximum
likelihood parameter values, while the 68\% confidence level (CL) is
given by $-2\Delta \ln \mathcal{L}=1$. Upper limits are quoted at 95\%
CL, i.e. given by $-2\Delta \ln \mathcal{L}= 2.71$, referring to
single-sided upper limits. The energy bins (i)-(iii) as defined in
Section~\ref{sec:data} were analyzed separately.

\begin{figure*}[t]
\centering
\includegraphics[width=\columnwidth]{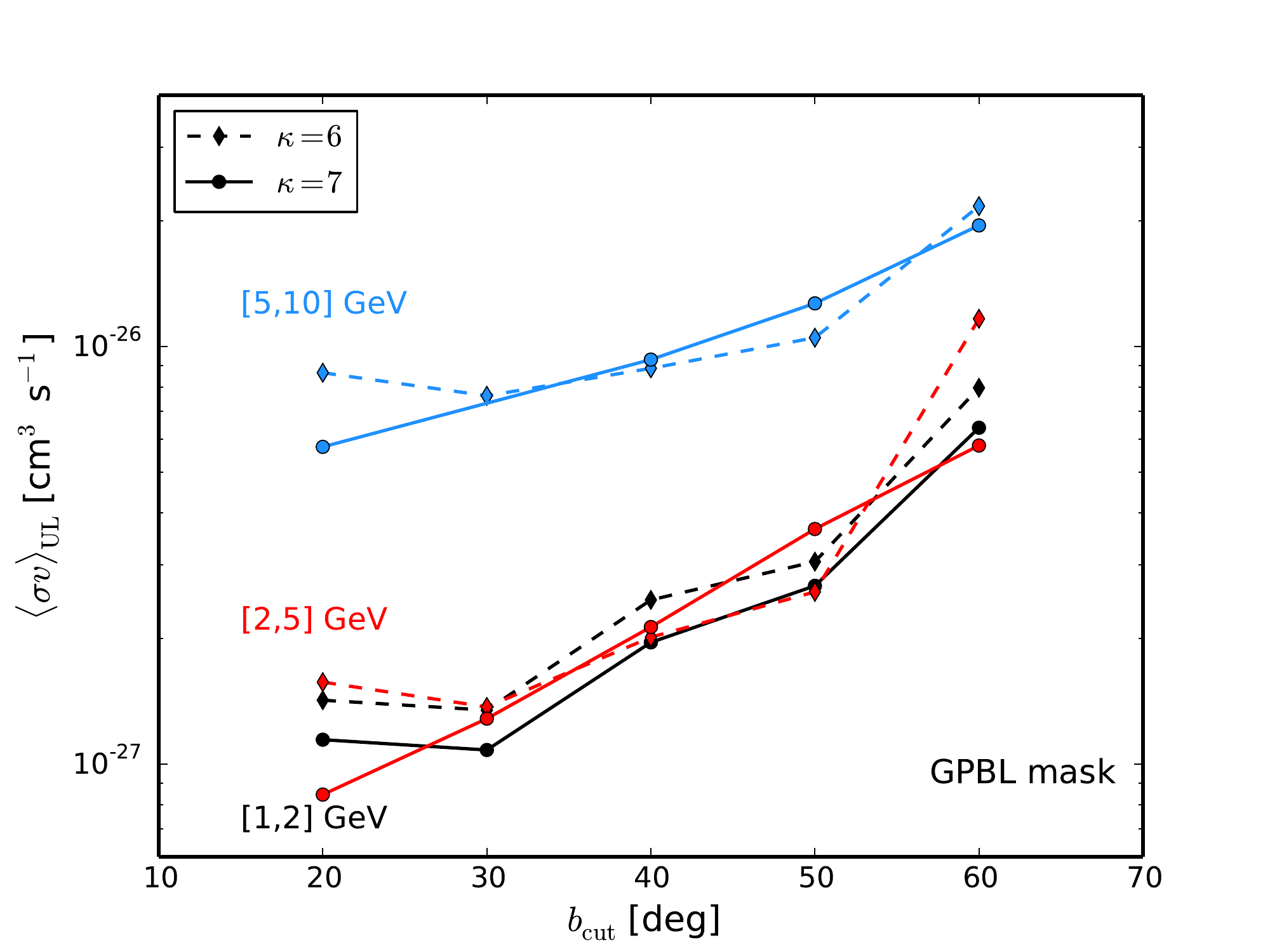}
\includegraphics[width=\columnwidth]{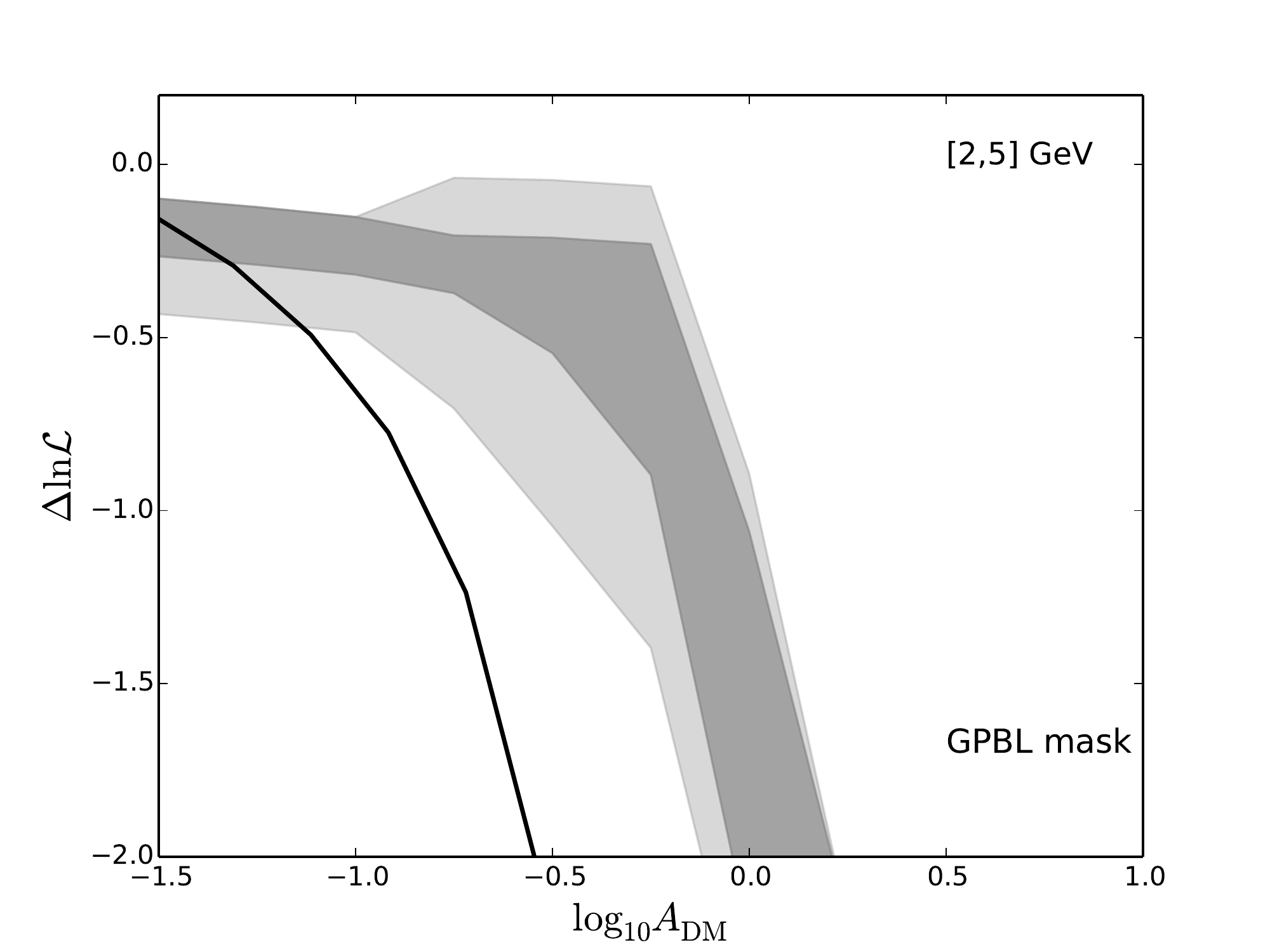}
\caption{\label{fig:bcut_HP} Left panel: Upper limits on the DM
  self-annihilation cross section $\langle \sigma v \rangle$ as
  function of the Galactic latitude cut $b_\mathrm{cut}$ and the pixel
  size. The analyzed ROI corresponds to the entire sky, excluding the
  GP region $|b| < b_\mathrm{cut}$, the Fermi Bubbles, and Galactic
  Loop~I (i.e. the GPBL mask). The limits refer to a DM mass of
  15\,GeV, and the $\tau^+\tau^-$ annihilation channel. Results are
  shown for all three energy bands considered. The DM halo has been
  modeled with the Einasto profile. The dashed lines (diamonds) depict
  results obtained by using a HEALPix grid with order $\kappa=6$,
  while the solid lines (circles) show limits obtained with
  $\kappa=7$.  Right panel: Statistical behavior of the ROI as
  analyzed in the left panel for $b_\mathrm{cut} = 30^\circ$ and
  $\kappa=7$.  The gray-shaded bands depict the 68\% (darkgray) and
  95\% (light-gray) confidence intervals derived from the statistical
  scatter of the $A_\mathrm{DM}$ profile likelihood function, as
  obtained from simulations of the gamma-ray sky by assuming
  $A_\mathrm{DM} = 0$. The solid black line shows the corresponding
  result obtained from the real flight data.}
\end{figure*}

Figures~\ref{fig:bcut_HP} and \ref{fig:nullhyp} depict the profile
likelihood functions for the $A_\mathrm{DM}$ parameter (together with
the statistical distributions expected for the null hypothesis), given
DM setups which are exemplary for the results discussed in the
following sections.

\subsection{\label{ssec:dNdS}Source-count Distribution $\mathrm{d}N/\mathrm{d}S$}
The $\mathrm{d}N/\mathrm{d}S$ distribution was parameterized with an
MBPL with three free consecutive breaks. Correspondingly, the MBPL
contributes 8 degrees of freedom in total, i.e. 4 power-law indexes
and a normalization constant in addition to the break positions.  The
fitted $\mathrm{d}N/\mathrm{d}S$ distribution was consistent with
measurements derived from the 3FGL \cite{2015ApJS..218...23A}
point-source catalog (see Z16a,b for details) in all analyses. As
discussed in Z16a,b, the $\mathrm{d}N/\mathrm{d}S$ fit obtained from
the 1pPDF extends previous catalog measurements to the regime of
faint, unresolved sources. All fits were sufficiently stable and
converged.

\subsection{\label{ssec:Adm} Region of Interest Optimization}
In order to produce statistically stable and robust results, the
analysis was optimized with respect to the choice of the ROI and the
choice of the pixel size.  ROI optimization was based on two main
aspects, addressed in the following paragraphs: (i) systematics
related to Galactic foreground emission and (ii) statistical validity.

The DM density distribution peaks in the center of the Galactic DM
halo, and thus data from the central regions of the Galaxy could have
significant impact on constraining a potential DM contribution. As
detailed in Section~\ref{ssec:IEMs}, it is however well known that the
modeling of the strong foreground emission from the GC region is
equipped with high systematic uncertainties that could significantly
affect the 1pPDF. We therefore chose to mask the GC and GP emissions
by excluding low Galactic latitudes $|b|<b_\mathrm{cut}$ from the
ROI. Further systematics could be introduced by potential mismodelings
of the Fermi Bubbles \cite{2010ApJ...724.1044S,2014ApJ...793...64A}
and Galactic Loop~I \cite{2009arXiv0912.3478C} that were masked as
well. This mask, as depicted in Fig.~\ref{fig:j_factor} by the gray
region, will be referred to as \textit{GPBL mask} in the remainder. We
optimized the analysis setup with regard to the choice of
$b_\mathrm{cut}$.

Moreover, past studies have demonstrated that fitting diffuse
templates using ROIs which cover a large fraction of the sky could
potentially lead to over-subtraction issues.  In particular, as
discussed in Refs.~\cite{Daylan:2014rsa, Calore:2014xka,
  Cohen:2016uyg, Linden:2016rcf}, this is connected to possible
mismodelings of backgrounds, namely the Galactic diffuse emission.  To
mitigate a possible over-subtraction of background models in our
analysis, we reduced and optimized the size of the ROI, following an
approach similar to Ref.~\cite{Cohen:2016uyg}.  In addition, we
performed simulations of the gamma-ray sky in order to challenge our
analysis setup against the null-hypothesis, i.e. assuming a gamma-ray
sky without any DM component.  Simulations were carried out such that
to resemble the actual gamma-ray data as closely as possible, see
Section~\ref{ssec:sim} for further details.  The actual flight data as
well as simulated realizations of the sky were then analyzed with the
1pPDF setup, choosing the region introduced above as initial
ROI. Subsequently, the ROI was systematically trimmed in longitude and
latitude, until the statistical behavior of the flight data met the
statistical expectation derived from data simulated for the
null-hypothesis $A_\mathrm{DM}=0$.  The trimming was symmetric in both
East-West or North-South directions, respectively.  We used the
benchmark IEM, $m_\mathrm{DM} = 15\,\mathrm{GeV}$, and annihilation
into $\tau^+\tau^-$ for all analyses related to the optimization
process. Details are discussed in the following.

As a first step, we try to identify the optimum value for the Galactic
latitude cut $b_\mathrm{cut}$, while investigating possible
systematics due to the choice of the HEALPix resolution.  For each
energy bin, the left panel of Fig.~\ref{fig:bcut_HP} shows upper
limits on $\langle \sigma v \rangle$ as a function of the Galactic
latitude cut $b_\mathrm{cut}$. The figure displays results referring
to the GPBL mask only, i.e. here we consider the almost entire
extragalactic gamma-ray sky. All limits are shown for two HEALPix
resolutions, i.e. $\kappa=6$ and $\kappa=7$. As demonstrated by the
figure, $b_\mathrm{cut}$ values equal or above $\sim\!30^\circ$ yield
considerably stable upper limits, with monotonically decreasing
sensitivity caused by decreasing event statistics for larger
$b_\mathrm{cut}$ values.  Larger ROIs, corresponding to smaller values
of $b_\mathrm{cut}$ below $30^\circ$ might instead be affected by the
stronger Galactic foreground morphology and should therefore be
disregarded. Upper limits derived for different pixel sizes are almost
equal for $b_\mathrm{cut} > 30^\circ$, with only slight differences
presumably originating from sampling effects or other small
systematics.  In the following, all the analyses were carried out with
HEALPix resolution $\kappa=7$.

As argued above, the statistical validity of the ROI was challenged
with simulations of the gamma-ray sky. The right panel of
Fig.~\ref{fig:bcut_HP} compares the statistical expectation for the
null-hypothesis with the analysis of the actual flight data for the
GPBL mask. It can be seen that the $A_\mathrm{DM}$ profile likelihood
function is significantly below the statistical scatter of the
simulations.

In order to find a statistically valid analysis region, the ROI was
subsequently shrunk in Galactic longitude $l$ and latitude $b$. Given
the large extent of the Galactic Loop~I structure in the Northern
hemisphere, here we focussed on the study of the Southern hemisphere,
where the ROI could be placed comparably closer to the GC. In
particular, we studied the following ROIs: (i) stripe-shaped, e.g. $l
\in [0,360]$\,deg, $b \in [-40,-30], [-50,-40],
\mathrm{and}\ [-60,-50]$\,deg, and (ii) box-shaped, e.g. $l \in
\{[0,80],[280,360]\}$\,deg, $b \in [-60,-40]$\,deg. All three energy
bands were considered separately. Given the systematic reduction of
the ROI size in $l$ and $b$, we found that real flight data match the
sensitivity expected from simulations for ROIs with longitudes
$|l|<90^\circ$ (centered on $l=0^\circ$) and $b \in [-60,-30]$\,deg.

\begin{figure*}[t]
\centering
\includegraphics[width=0.32\textwidth]{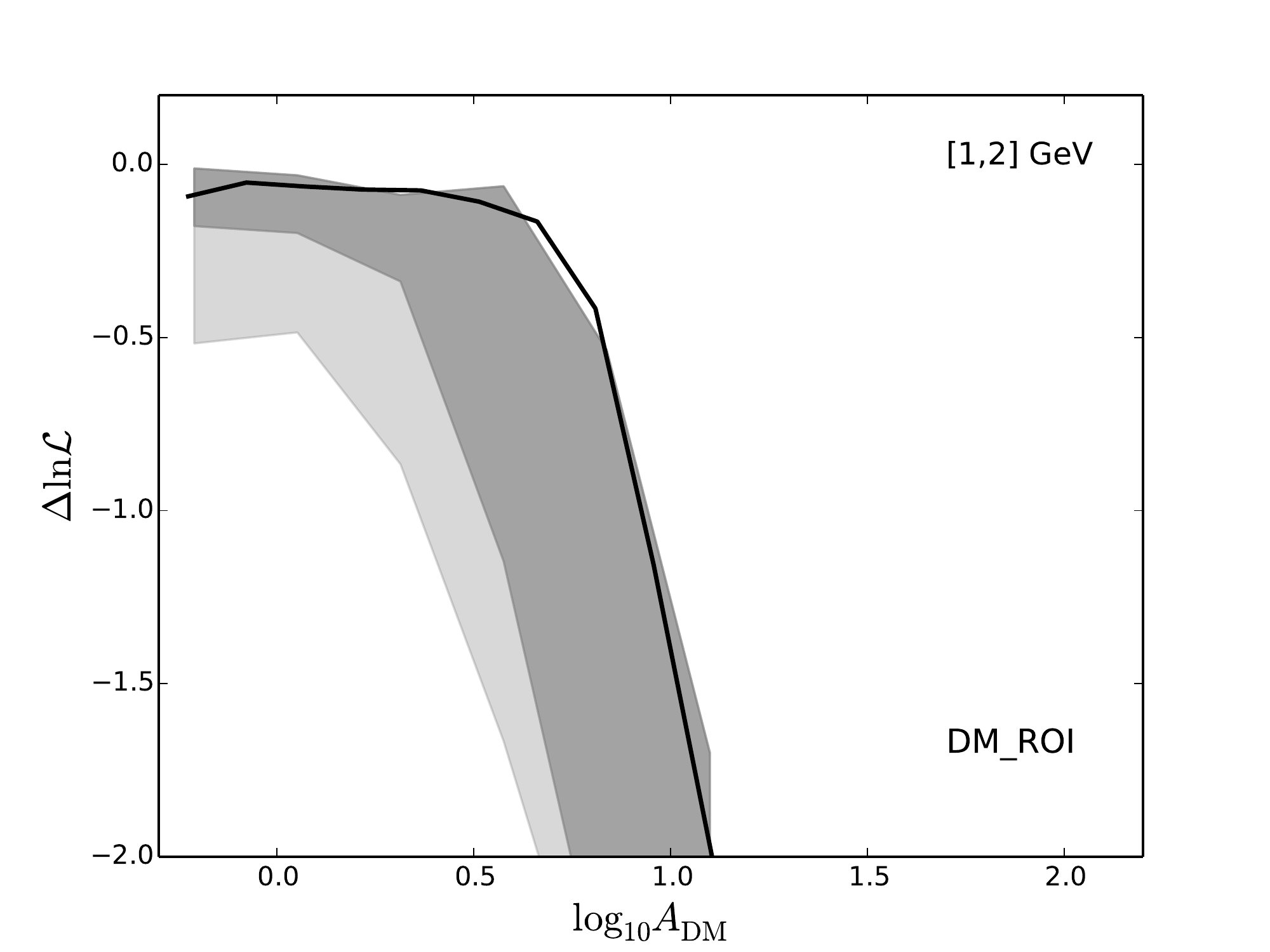}
\includegraphics[width=0.32\textwidth]{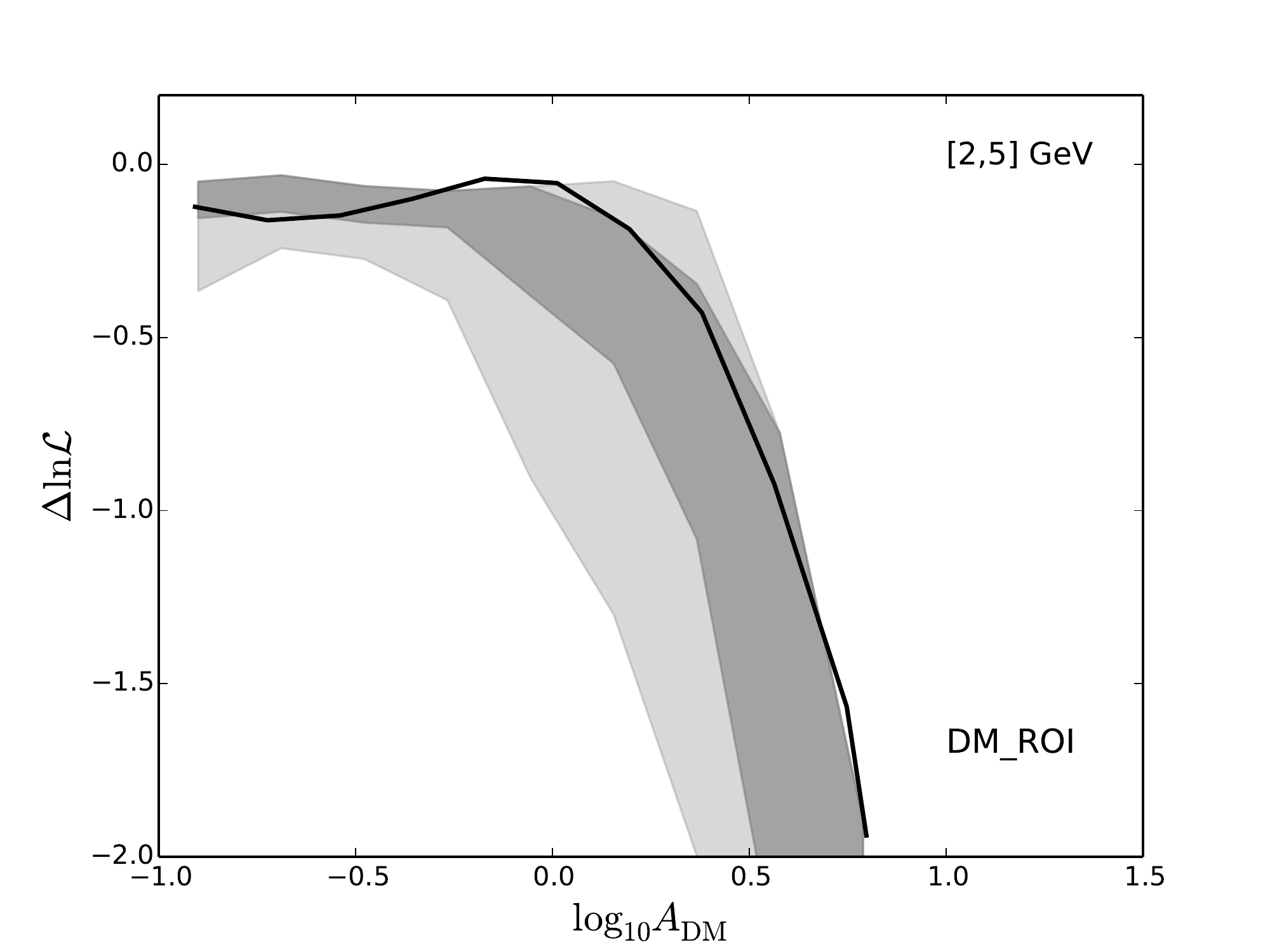}
\includegraphics[width=0.32\textwidth]{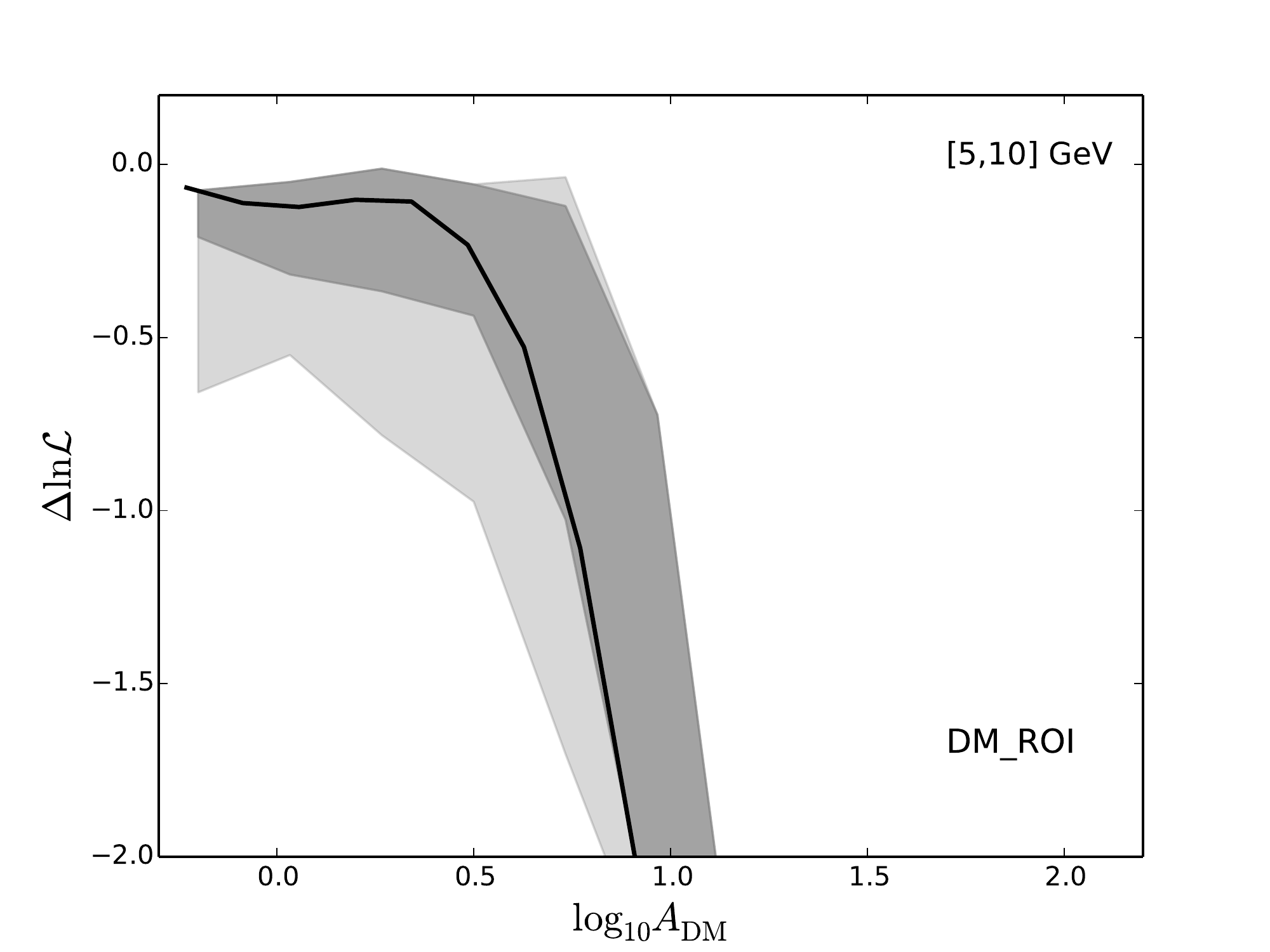}
\caption{\label{fig:nullhyp} Comparison of the actual flight data with
  the statistical expectation for the null-hypothesis as derived from
  simulations.  The DM\_ROI (see Fig.~\ref{fig:ROI}) has been
  considered in the three energy bands 1.04--1.99\,GeV (left panel),
  1.99--5.0\,GeV (middle panel), and 5.0--10.4\,GeV (right panel). The
  analysis setup refers to a DM mass of 15\,GeV, and the
  $\tau^+\tau^-$ annihilation channel. The gray-shaded bands depict
  the 68\% (darkgray) and 95\% (light-gray) confidence intervals
  derived from the statistical scatter of the $A_\mathrm{DM}$ profile
  likelihood functions, as obtained from simulations of the gamma-ray
  sky by assuming $A_\mathrm{DM} = 0$. The solid black line shows the
  corresponding result obtained from the actual data.}
\end{figure*}

\begin{figure}
\centering
\includegraphics[width=\columnwidth]{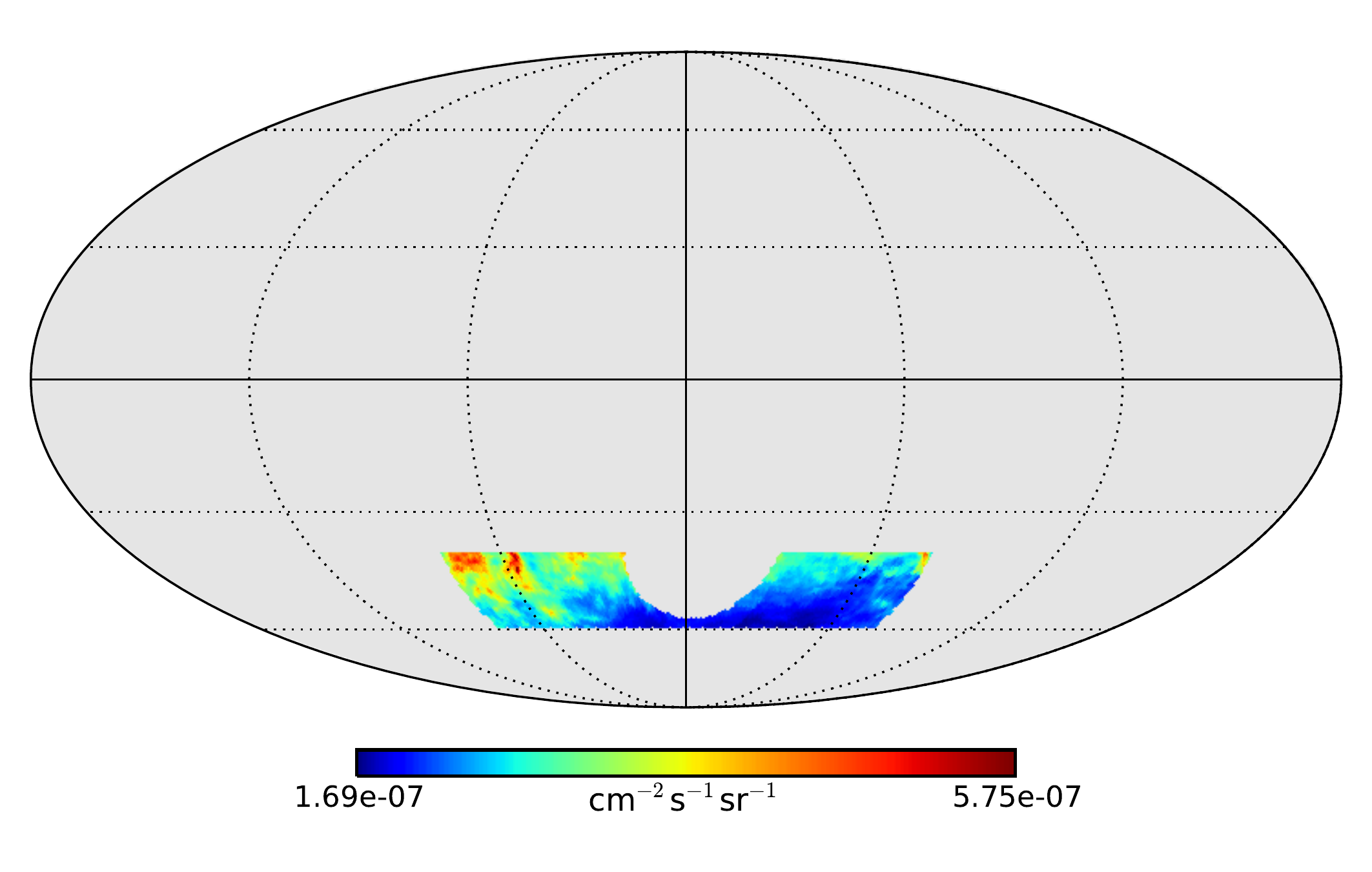}
\caption{\label{fig:ROI}The DM\_ROI considered in this analysis. The
  Mollweide projection depicts the benchmark IEM template in Galactic
  coordinates, centered on the GC. The map shows the integral flux
  $F_\mathrm{gal}$ between 1.99 and 5.0\,GeV. The area outside the
  DM\_ROI has been masked in gray.}
\end{figure}

Figure~\ref{fig:nullhyp} compares the statistical behavior of the
simulations with actual flight data for the DM\_ROI defined by $l \in
\{[0,80],[280,360]\}$\,deg and $b \in [-60,-40]$\,deg, which we chose
as benchmark ROI due to stability and robustness.  It can be seen that
the statistical behavior of the flight data is well consistent with
the expectation. Larger ROIs (within the allowed ranges as given
above) may slightly improve sensitivity by a factor of $<2$. The
chosen DM\_ROI is shown in Fig.~\ref{fig:ROI}, demonstrating the
influence of the benchmark IEM in the ROI.

\subsection{\label{ssec:sim} Simulations of the Fermi Sky}
Realistic Monte Carlo simulations of the gamma-ray sky were produced
using the \texttt{gtobsim} utility of the Fermi Science Tools v10r0p5,
as similarly done in \cite{2016ApJS..225...18Z}. The considered time
interval, as well as event class and energy range match the selection
done for the real flight data discussed in Section~\ref{sec:data}.

Three components enter the simulated counts map: point sources,
Galactic foreground, and diffuse isotropic background emission.  The
flux distribution of point sources $\mathrm{d}N/\mathrm{d}S$ was taken
as the best fit to the real data (see Section~\ref{ssec:dNdS}).  For
each simulation a list of point sources was produced with a Monte
Carlo simulation, with fluxes following the chosen
$\mathrm{d}N/\mathrm{d}S$ distribution and random positions across the
sky.  The flux spectrum of simulated sources was taken to be a power
law, where the photon index for each source was drawn from a Gaussian
distribution with mean $\Gamma=2.4$ and standard deviation
$\sigma_{\Gamma}=0.4$.  Point sources were simulated down to fluxes of
$S_{\rm min}=10^{-12}$\,cm$^{-2}$\,s$^{-1}$.  The official diffuse
Galactic emission template named \texttt{gll\_iem\_v06.fits} was
used. For the isotropic emission, we used the recommended spectral
template corresponding to our data selection,
\texttt{iso\_P8R2\_ULTRACLEANVETO\_V6\_PSF3\_v06.txt}.  The
normalization of the isotropic emission was chosen to match the
integral flux $F_{\rm iso}$ observed in real data (see
Tab.~\ref{tab:composition}).  An effective PSF correction was computed
according to the simulation properties.  The resulting mock data maps
were then analyzed with the same analysis chain as used for the real
data, see Section~\ref{sec:data}.

\section{\label{sec:results}Results}
\begin{figure*}
\centering
\includegraphics[width=\columnwidth]{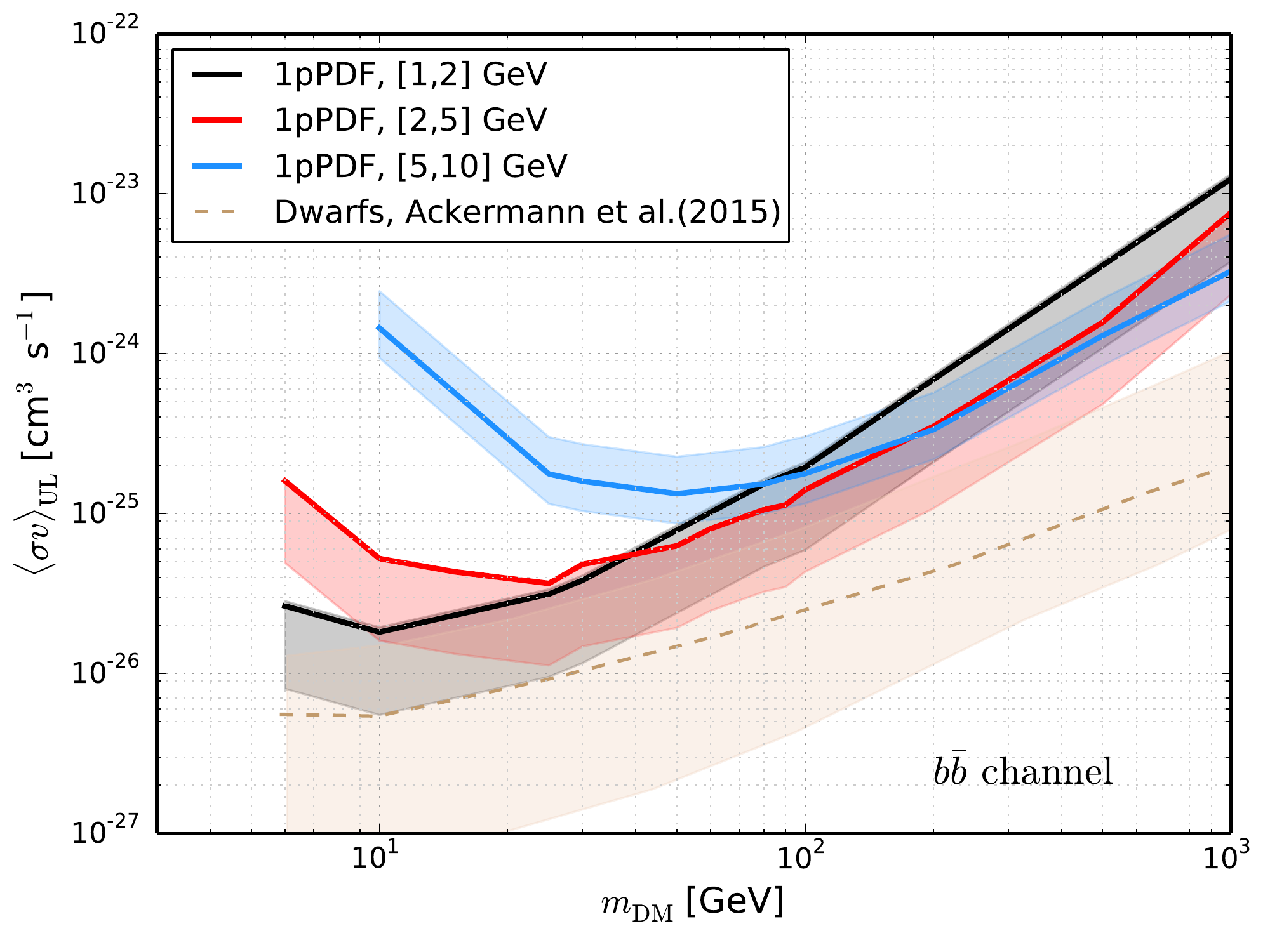}
\includegraphics[width=\columnwidth]{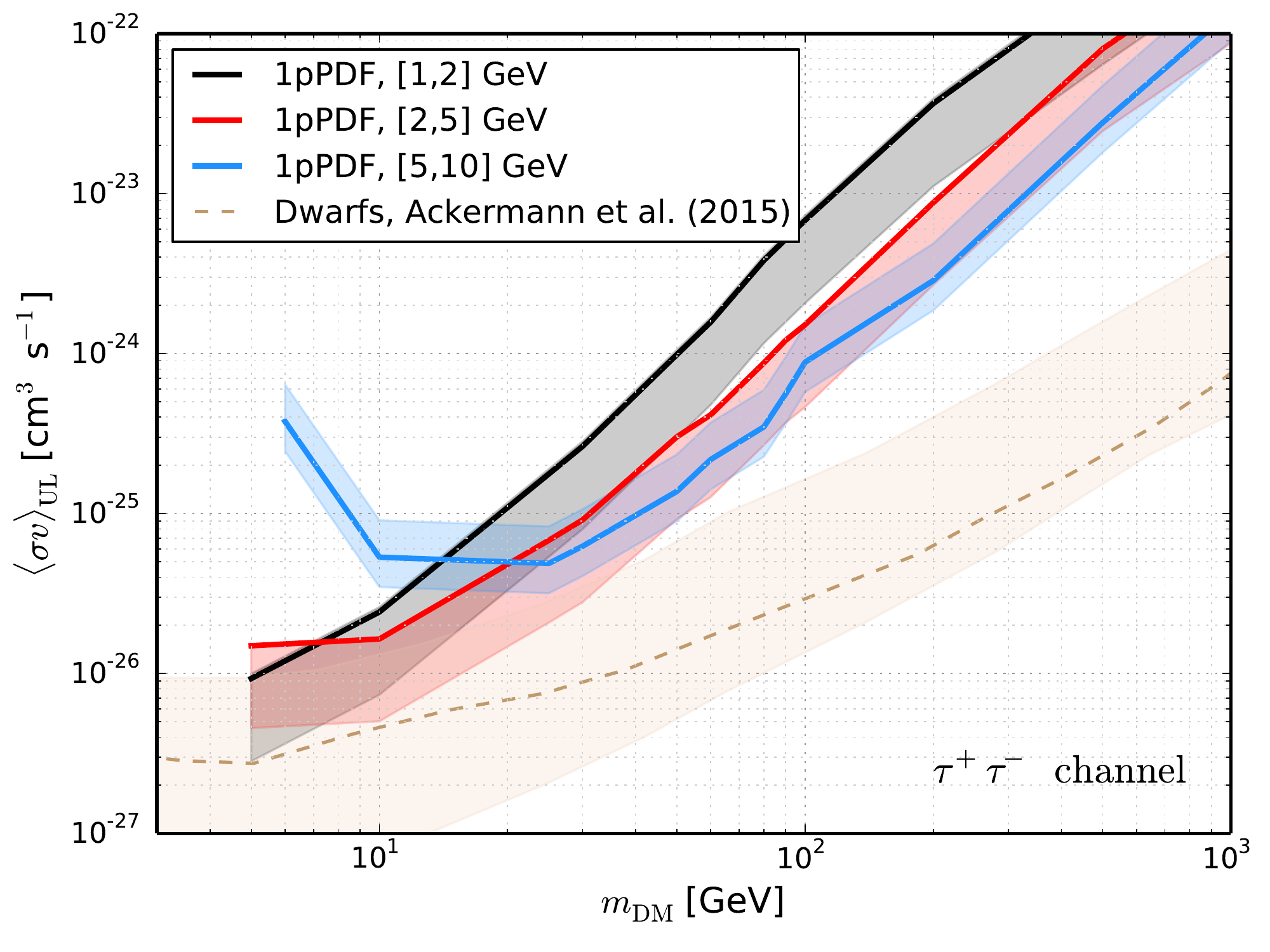}
\caption{\label{fig:ul_results} Upper limits (95\% CL) on the DM
  self-annihilation cross section $\langle \sigma v \rangle$ as a
  function of the DM particle mass $m_\mathrm{DM}$, as obtained with
  the 1pPDF analysis using the DM\_ROI for 8-year Fermi-LAT data ({\tt
    Pass 8}).  The DM halo of the Galaxy was assumed to follow an
  Einasto profile. Upper limits are given for separate analyses of the
  (i) 1.04--1.99\,GeV (black solid line), (ii) 1.99--5.0\,GeV (red
  solid line), and (iii) 5.0--10.4\,GeV (blue solid line) energy
  bins. The shaded bands reflect the expected sensitivity (95\%
  confidence level) as derived from simulations. The left (right)
  panel shows upper limits for total annihilation into $b\overline{b}$
  ($\tau^+\tau^-$) final states. The limits are compared to recent
  limits obtained from the observation of dwarf spheroidal galaxies,
  see Ref.~\cite{2015PhRvL.115w1301A} (orange dashed line and shaded
  region).}
\end{figure*}

Figure~\ref{fig:ul_results} shows the upper limits on the
self-annihilation cross section $\langle \sigma v \rangle$ of DM
particles (such as WIMPs) annihilating to $b\overline{b}$ (left panel)
and $\tau^+\tau^-$ (right panel) final states, obtained with the 1pPDF
setup as developed above. The upper limits were derived using the
benchmark IEM, for DM particles with masses between 5\,GeV and
1\,TeV. The 1pPDF analysis was performed on the DM\_ROI data.  The
three adjacent energy bins (i)~1.04--1.99\,GeV, (ii)~1.99--5.0\,GeV,
and (iii)~5.0--10.4\,GeV were considered to be independent from each
other, yielding three different results for each annihilation
channel. Using the benchmark IEM, no evidence was found for the
additional DM component to significantly improve the quality of the
fit, corresponding to an open, single-sided profile likelihood curve
for the $A_\mathrm{DM}$ parameter (cf. Fig.~\ref{fig:nullhyp}). The
different shapes of the curves for $b\overline{b}$ and $\tau^+\tau^-$
final states originate from the different gamma-ray emission spectra
$\mathrm{d}N_f/\mathrm{d}E$, where in particular annihilation to
$b\overline{b}$ yields softer gamma-ray spectra than annihilation to
$\tau^+\tau^-$.

For all upper bounds, we also show the 95\% confidence level expected
sensitivity derived from the simulations, as described in the previous
section. Figure~\ref{fig:ul_results} compares the 1pPDF results to
upper limits obtained from the stacking of several dwarf spheroidal
galaxies (dSphs) \cite{2015PhRvL.115w1301A}. We find that the
sensitivity reach of the 1pPDF analysis is comparable with the
analysis of dSphs, in particular for dark matter masses below 100 GeV
(depending on the annihilation channel). Note that the different shape
of the 1pPDF limits with respect to the dSph limits is owed to the
integration over different energy intervals.

The 1pPDF fit decomposes the gamma-ray sky according to the modeling
discussed in Section~\ref{ssec:IEMs}. For each energy bin,
Tab.~\ref{tab:composition} lists the fractional contributions from the
three main components to the total integral flux $F_\mathrm{tot}$,
i.e. from point sources ($q_{\rm ps}$), from the benchmark IEM
($q_{\rm gal}$), and from the diffuse isotropic background component
($q_{\rm iso}$). Given the lacking significance for a possible DM
component, here its contribution is assumed to be negligible.  The
quantities in Tab.~\ref{tab:composition} refer to the 1pPDF analysis
using GPBL mask. Very similar results are found using our final
DM\_ROI, with larger uncertainties due to smaller statistics.

\begin{table}[t]
\caption{\label{tab:composition}Composition of the gamma-ray sky for
  $|b|\geq 30^\circ$. The quantities $q_{\rm ps}$, $q_{\rm gal}$, and
  $q_{\rm iso}$ denote the fractional contribution from the
  corresponding component to the integral map flux
  $F_\mathrm{tot}$. The total flux $F_\mathrm{tot}$ is given in units
  of $10^{-7}\,\mathrm{cm}^{-2}\mathrm{s}^{-1}\mathrm{sr}^{-1}$.}
\begin{ruledtabular}
\begin{tabular}{lccc}
Component & 1.04$-$1.99\,GeV  & 1.99$-$5.0\,GeV & 5.0$-$10.4\,GeV \\
\hline
Sources ($q_{\rm ps}$)  &  $0.28^{+0.03}_{-0.03}$ &	$0.21^{+0.03}_{-0.02}$ &	$0.21^{+0.04}_{-0.03}$  \\
IEM ($q_{\rm gal}$) &	$0.714^{+0.007}_{-0.005}$&	$0.675^{+0.008}_{-0.011}$ &	$0.548^{+0.019}_{-0.018}$  \\
Isotropic ($q_{\rm iso}$) & $0.03^{+0.02}_{-0.01}$ &	$0.12^{+0.03}_{-0.04}$ &	$0.24^{+0.05}_{-0.05}$ \\
\hline
$F_\mathrm{tot}$ & $7.828^{+0.016}_{-0.016}$  & $3.875^{+0.111}_{-0.111}$ &  $0.951^{+0.005}_{-0.005}$  \\
\end{tabular}
\end{ruledtabular}
\end{table}

The upper limits presented in Fig. \ref{fig:ul_results} were obtained
for the benchmark, official \emph{Fermi} IEM. Possible degeneracies
with the IEM as a single component were incorporated by means of the
normalization parameter $A_\mathrm{gal}$. As such, the results
presented in the figure reflect statistically valid upper limits under
the assumption that systematic uncertainties of the IEM and its
constituents are small as compared to statistical
uncertainties. However, degeneracies between the DM component with
particular IEM constituents, such as IC emission, remain possible.

\begin{figure*}
\centering
\includegraphics[width=\columnwidth]{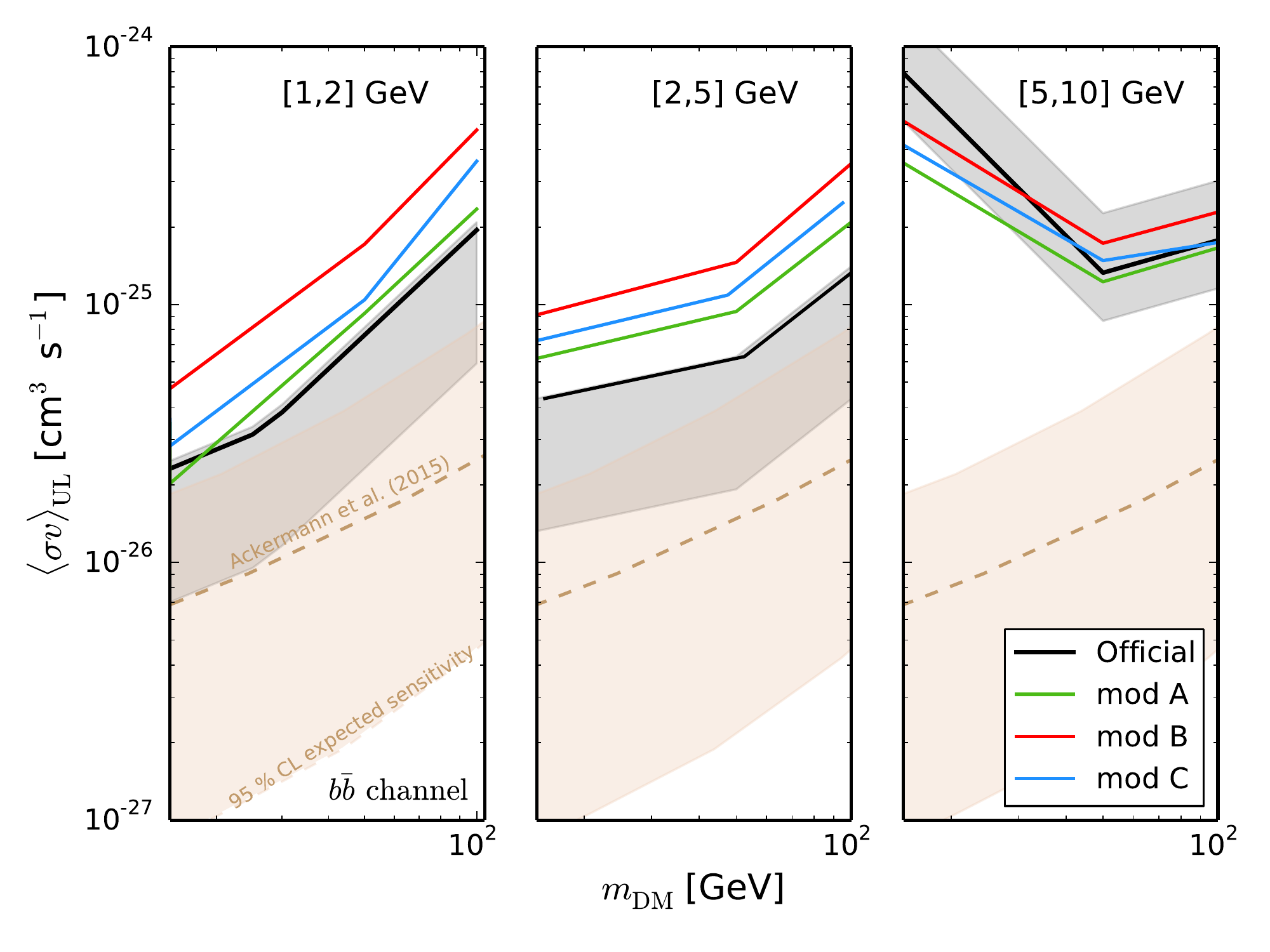}
\includegraphics[width=\columnwidth]{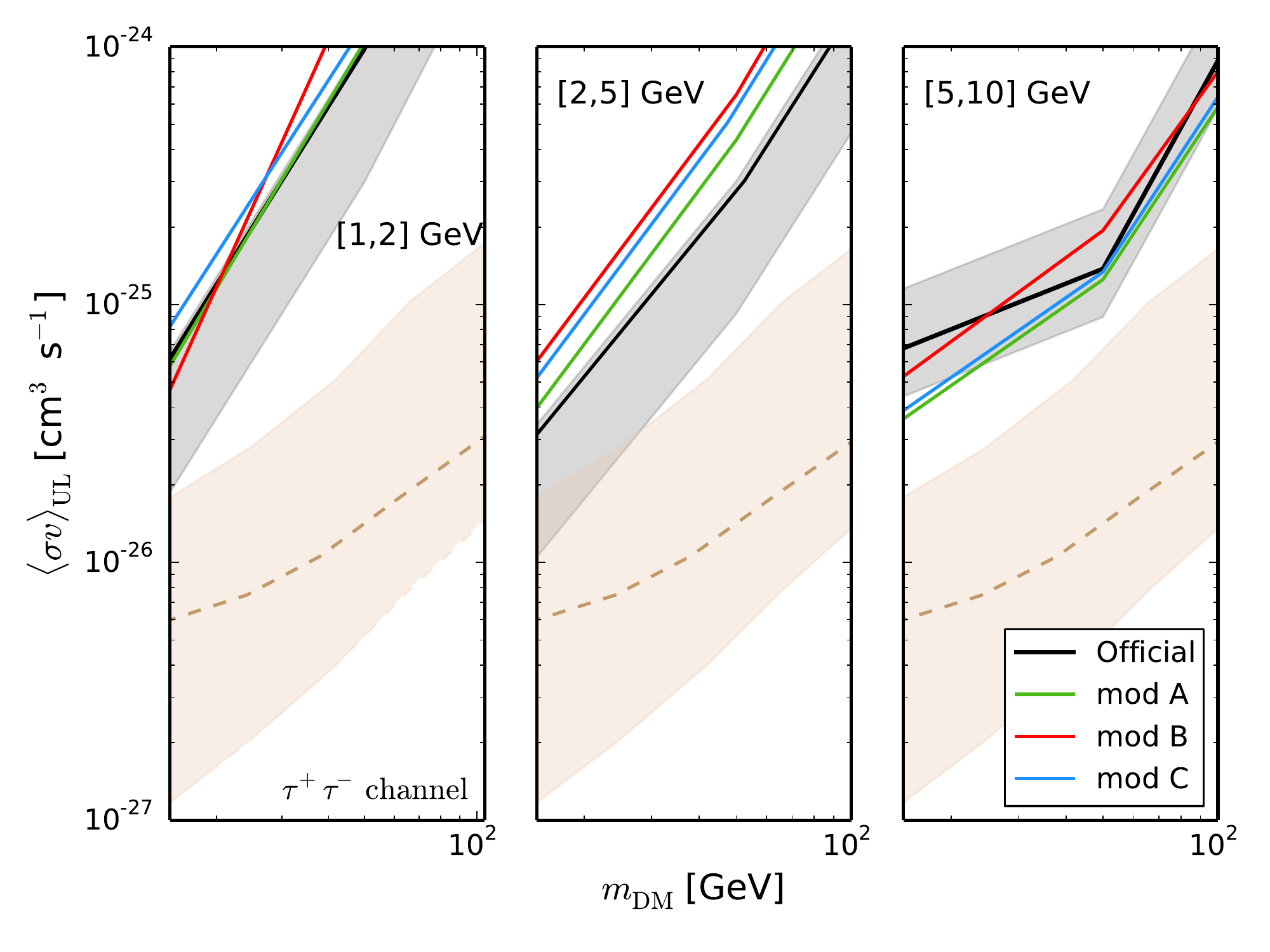}
\caption{\label{fig:ul_modABCresults} Upper limits (95\% CL) on the DM
  self-annihilation cross section $\langle \sigma v \rangle$ for
  $b\overline{b}$ (left panel) and $\tau^+\tau^-$ (right panel) final
  states and $m_\mathrm{DM} = 15, 50, 100$~GeV using the DM\_ROI for
  8-year Fermi-LAT data ({\tt Pass 8}), obtained assuming the
  benchmark IEM (black solid line), model A (green solid line), model
  B (red solid line), and model C (blue solid line) as discussed in
  Section~\ref{ssec:IEMs}. The DM halo of the Galaxy was assumed to
  follow an Einasto profile. Upper limits are given for the three
  energy bins (i) 1.04--1.99\,GeV (left panel), (ii) 1.99--5.0\,GeV
  (middle panel), and (iii) 5.0--10.4\,GeV (right panel). The limits
  are compared to the limits obtained from the observation of dwarf
  spheroidal galaxies, see Ref.~\cite{2015PhRvL.115w1301A} (orange
  dashed line). For illustrative purposes, the yellow band depicts the
  95\% quantile of the median expected sensitivity of the dwarf
  spheroidal galaxy analysis.}
\end{figure*}

To estimate the scatter of the upper limits with respect to the
diffuse Galactic foreground emission, Fig.~\ref{fig:ul_modABCresults}
compares the results obtained previously to upper limits derived for
the selection of three other IEMs. The IEMs considered here were
selected to bracket plausible Galactic foreground emission
scenarios. We chose models A, B, C as discussed in
Section~\ref{ssec:IEMs}. The figure depicts upper limits for DM
particle masses $15, 50$, and 100\,GeV, considering annihilation into
$b\overline{b}$ and $\tau^+\tau^-$. We find that the upper limits
obtained for model B are almost always the least constraining, because
its IC emission component for $|b|>30^\circ$ is less prominent than in
model A and C, thus leaving room for a larger DM contribution. The
amplitude of the scatter due to the different IEMs is about a factor
2-3, depending on the energy bin, and is therefore comparable to the
band of the expected sensitivity inherent to our analysis
method. These upper bounds on $\langle \sigma v \rangle$ are compared
to the limits obtained from the observation of dwarf spheroidal
galaxies, see Ref.~\cite{2015PhRvL.115w1301A}.

\section{\label{sec:conclusions}Conclusions} 
It has recently been shown (see Z16a,b) that statistical properties of
the {\it Fermi}-LAT photon counts map can be used to measure the
composition of the gamma-ray sky at high latitudes, determining
$\mathrm{d}N/\mathrm{d}S$ down to fluxes about one order of magnitude
lower than current catalog detection limits. The high latitude
gamma-ray sky is modeled with at least three components, represented
by an isotropic distribution of point sources, a diffuse component of
Galactic foreground emission, and diffuse isotropic background. In
this paper, we have extended the photon count statistics 1pPDF method
developed in Z16a,b to a further component of the high-latitude sky,
given by Galactic DM distributed in a typical smooth halo. We have
employed the 1pPDF method to derive upper bounds on the possible
contribution from halo DM in terms of the self-annihilation cross
section $\langle \sigma v \rangle$, for DM masses spanning the GeV to
TeV range.

We find that the 1pPDF method applied to eight years of \texttt{Pass
  8} {\it Fermi}-LAT data at high latitudes has the sensitivity for
assessing the possible gamma-ray contribution of Galactic DM
annihilating into $b\overline{b}$ or $\tau^+\tau^-$ final
states. However, we find that the analysis can be affected by
over-subtraction of the background IEM when the ROI covers a
significant portion of the sky. We have found that a reliable ROI for
the DM analysis is a small box of the sky located in the southern
hemisphere (DM\_ROI). Given the official {\it Fermi}-LAT interstellar
emission model, the upper bounds obtained for $\langle \sigma v
\rangle$ are comparable to constraints from the stacking analysis of
several dwarf spheroidal galaxies. The analysis comprises three
adjacent bins in photon energy, spanning from 1 to 10\,GeV. The three
bins are increasingly relevant with increasing $m_{\rm DM}$.

Our results have been verified against sky simulations realized with
the Fermi science tools, without DM templates. The 1pPDF analysis
provides results coherent with the expected sensitivity derived from
the simulations, once the ROI is properly optimized.  Pixelizing the
sky map with different resolutions provides stable results.

Eventually, we have repeated our analysis for three additional IEM
templates. The modeling of the Galactic diffuse emission has a
non-negligible systematic impact, given that the upper bound on
$\langle \sigma v \rangle$ can vary by a factor of a few, depending on
the energy bin and the DM mass. Using the {\it Fermi}-LAT official
template provides stronger bounds than the models including smaller IC
emission at high latitudes. This is expected, given the possible
degeneracy between IC and halo DM maps.

We have demonstrated that the method of 1-point photon count
statistics, when applied to eight years of {\it Fermi}-LAT data, has
the sensitivity for assessing a possible DM contribution to the high
latitude sky down to DM self-annihilation cross section $\langle
\sigma v \rangle$ comparable to the ones bound by the currently most
powerful, complementary methods.

\begin{acknowledgments}
We warmly thank A.~Cuoco, N.~Fornengo, D.~Malyshev, and M.~Regis, and
SM warmly thanks M.~Di~Mauro and E.~Charles, for useful discussions
and insights. We thank the anonymous referee for their thorough check
of our results and very helpful comments. SM gratefully acknowledges
support by the NASA {\it Fermi} Guest Investigator Program 2016
through the {\it Fermi} one-year Program N.\,91245 (P.I. M.~Di~Mauro),
support by the Academy of Science of Torino through the {\it Angiola
  Agostinelli Gili} 2016 scholarship, and the KIPAC institute at SLAC
for the kind hospitality, where a part of this project was
completed. HSZ gratefully acknowledges the Istituto Nazionale di
Fisica Nucleare (INFN) for a post-doctoral fellowship in theoretical
physics on "Astroparticle, Dark Matter and Neutrino Physics", awarded
under the INFN Fellowship Programme 2015.
\end{acknowledgments}

\bibliography{ms}

\begin{thebibliography}{66}%
\makeatletter
\providecommand \@ifxundefined [1]{%
 \@ifx{#1\undefined}
}%
\providecommand \@ifnum [1]{%
 \ifnum #1\expandafter \@firstoftwo
 \else \expandafter \@secondoftwo
 \fi
}%
\providecommand \@ifx [1]{%
 \ifx #1\expandafter \@firstoftwo
 \else \expandafter \@secondoftwo
 \fi
}%
\providecommand \natexlab [1]{#1}%
\providecommand \enquote  [1]{``#1''}%
\providecommand \bibnamefont  [1]{#1}%
\providecommand \bibfnamefont [1]{#1}%
\providecommand \citenamefont [1]{#1}%
\providecommand \href@noop [0]{\@secondoftwo}%
\providecommand \href [0]{\begingroup \@sanitize@url \@href}%
\providecommand \@href[1]{\@@startlink{#1}\@@href}%
\providecommand \@@href[1]{\endgroup#1\@@endlink}%
\providecommand \@sanitize@url [0]{\catcode `\\12\catcode `\$12\catcode
  `\&12\catcode `\#12\catcode `\^12\catcode `\_12\catcode `\%12\relax}%
\providecommand \@@startlink[1]{}%
\providecommand \@@endlink[0]{}%
\providecommand \url  [0]{\begingroup\@sanitize@url \@url }%
\providecommand \@url [1]{\endgroup\@href {#1}{\urlprefix }}%
\providecommand \urlprefix  [0]{URL }%
\providecommand \Eprint [0]{\href }%
\providecommand \doibase [0]{http://dx.doi.org/}%
\providecommand \selectlanguage [0]{\@gobble}%
\providecommand \bibinfo  [0]{\@secondoftwo}%
\providecommand \bibfield  [0]{\@secondoftwo}%
\providecommand \translation [1]{[#1]}%
\providecommand \BibitemOpen [0]{}%
\providecommand \bibitemStop [0]{}%
\providecommand \bibitemNoStop [0]{.\EOS\space}%
\providecommand \EOS [0]{\spacefactor3000\relax}%
\providecommand \BibitemShut  [1]{\csname bibitem#1\endcsname}%
\let\auto@bib@innerbib\@empty
\bibitem [{\citenamefont {{Atwood}}\ \emph {et~al.}(2009)\citenamefont
  {{Atwood}}, \citenamefont {{Abdo}}, \citenamefont {{Ackermann}},
  \citenamefont {{Althouse}}, \citenamefont {{Anderson}} \emph
  {et~al.}}]{2009ApJ...697.1071A}%
  \BibitemOpen
  \bibfield  {author} {\bibinfo {author} {\bibfnamefont {W.~B.}\ \bibnamefont
  {{Atwood}}}, \bibinfo {author} {\bibfnamefont {A.~A.}\ \bibnamefont
  {{Abdo}}}, \bibinfo {author} {\bibfnamefont {M.}~\bibnamefont {{Ackermann}}},
  \bibinfo {author} {\bibfnamefont {W.}~\bibnamefont {{Althouse}}}, \bibinfo
  {author} {\bibfnamefont {B.}~\bibnamefont {{Anderson}}},  \emph {et~al.},\
  }\href {\doibase 10.1088/0004-637X/697/2/1071} {\bibfield  {journal}
  {\bibinfo  {journal} {\apj}\ }\textbf {\bibinfo {volume} {697}},\ \bibinfo
  {pages} {1071} (\bibinfo {year} {2009})},\ \Eprint
  {http://arxiv.org/abs/0902.1089} {arXiv:0902.1089 [astro-ph.IM]} \BibitemShut
  {NoStop}%
\bibitem [{\citenamefont {{Ackermann}}\ \emph
  {et~al.}(2012{\natexlab{a}})\citenamefont {{Ackermann}}, \citenamefont
  {{Ajello}}, \citenamefont {{Albert}}, \citenamefont {{Allafort}},
  \citenamefont {{Atwood}} \emph {et~al.}}]{2012ApJS..203....4A}%
  \BibitemOpen
  \bibfield  {author} {\bibinfo {author} {\bibfnamefont {M.}~\bibnamefont
  {{Ackermann}}}, \bibinfo {author} {\bibfnamefont {M.}~\bibnamefont
  {{Ajello}}}, \bibinfo {author} {\bibfnamefont {A.}~\bibnamefont {{Albert}}},
  \bibinfo {author} {\bibfnamefont {A.}~\bibnamefont {{Allafort}}}, \bibinfo
  {author} {\bibfnamefont {W.~B.}\ \bibnamefont {{Atwood}}},  \emph {et~al.},\
  }\href {\doibase 10.1088/0067-0049/203/1/4} {\bibfield  {journal} {\bibinfo
  {journal} {\apjs}\ }\textbf {\bibinfo {volume} {203}},\ \bibinfo {eid} {4}
  (\bibinfo {year} {2012}{\natexlab{a}})},\ \Eprint
  {http://arxiv.org/abs/1206.1896} {arXiv:1206.1896 [astro-ph.IM]} \BibitemShut
  {NoStop}%
\bibitem [{\citenamefont {{Fornasa}}\ and\ \citenamefont
  {{S{\'a}nchez-Conde}}(2015)}]{2015PhR...598....1F}%
  \BibitemOpen
  \bibfield  {author} {\bibinfo {author} {\bibfnamefont {M.}~\bibnamefont
  {{Fornasa}}}\ and\ \bibinfo {author} {\bibfnamefont {M.~A.}\ \bibnamefont
  {{S{\'a}nchez-Conde}}},\ }\href {\doibase 10.1016/j.physrep.2015.09.002}
  {\bibfield  {journal} {\bibinfo  {journal} {\physrep}\ }\textbf {\bibinfo
  {volume} {598}},\ \bibinfo {pages} {1} (\bibinfo {year} {2015})}\BibitemShut
  {NoStop}%
\bibitem [{\citenamefont {{Ackermann}}\ \emph {et~al.}(2015)\citenamefont
  {{Ackermann}}, \citenamefont {{Ajello}}, \citenamefont {{Albert}},
  \citenamefont {{Atwood}}, \citenamefont {{Baldini}} \emph
  {et~al.}}]{2015ApJ...799...86A}%
  \BibitemOpen
  \bibfield  {author} {\bibinfo {author} {\bibfnamefont {M.}~\bibnamefont
  {{Ackermann}}}, \bibinfo {author} {\bibfnamefont {M.}~\bibnamefont
  {{Ajello}}}, \bibinfo {author} {\bibfnamefont {A.}~\bibnamefont {{Albert}}},
  \bibinfo {author} {\bibfnamefont {W.~B.}\ \bibnamefont {{Atwood}}}, \bibinfo
  {author} {\bibfnamefont {L.}~\bibnamefont {{Baldini}}},  \emph {et~al.},\
  }\href {\doibase 10.1088/0004-637X/799/1/86} {\bibfield  {journal} {\bibinfo
  {journal} {\apj}\ }\textbf {\bibinfo {volume} {799}},\ \bibinfo {eid} {86}
  (\bibinfo {year} {2015})}\BibitemShut {NoStop}%
\bibitem [{\citenamefont {{Inoue}}\ and\ \citenamefont
  {{Totani}}(2009)}]{2009ApJ...702..523I}%
  \BibitemOpen
  \bibfield  {author} {\bibinfo {author} {\bibfnamefont {Y.}~\bibnamefont
  {{Inoue}}}\ and\ \bibinfo {author} {\bibfnamefont {T.}~\bibnamefont
  {{Totani}}},\ }\href {\doibase 10.1088/0004-637X/702/1/523} {\bibfield
  {journal} {\bibinfo  {journal} {\apj}\ }\textbf {\bibinfo {volume} {702}},\
  \bibinfo {eid} {523-536} (\bibinfo {year} {2009})},\ \Eprint
  {http://arxiv.org/abs/0810.3580} {arXiv:0810.3580} \BibitemShut {NoStop}%
\bibitem [{\citenamefont {{Ackermann}}\ \emph {et~al.}(2011)\citenamefont
  {{Ackermann}}, \citenamefont {{Ajello}}, \citenamefont {{Allafort}},
  \citenamefont {{Antolini}}, \citenamefont {{Atwood}} \emph
  {et~al.}}]{2011ApJ...743..171A}%
  \BibitemOpen
  \bibfield  {author} {\bibinfo {author} {\bibfnamefont {M.}~\bibnamefont
  {{Ackermann}}}, \bibinfo {author} {\bibfnamefont {M.}~\bibnamefont
  {{Ajello}}}, \bibinfo {author} {\bibfnamefont {A.}~\bibnamefont
  {{Allafort}}}, \bibinfo {author} {\bibfnamefont {E.}~\bibnamefont
  {{Antolini}}}, \bibinfo {author} {\bibfnamefont {W.~B.}\ \bibnamefont
  {{Atwood}}},  \emph {et~al.},\ }\href {\doibase 10.1088/0004-637X/743/2/171}
  {\bibfield  {journal} {\bibinfo  {journal} {\apj}\ }\textbf {\bibinfo
  {volume} {743}},\ \bibinfo {eid} {171} (\bibinfo {year} {2011})},\ \Eprint
  {http://arxiv.org/abs/1108.1420} {arXiv:1108.1420 [astro-ph.HE]} \BibitemShut
  {NoStop}%
\bibitem [{\citenamefont {{Abazajian}}\ \emph {et~al.}(2011)\citenamefont
  {{Abazajian}}, \citenamefont {{Blanchet}},\ and\ \citenamefont
  {{Harding}}}]{2011PhRvD..84j3007A}%
  \BibitemOpen
  \bibfield  {author} {\bibinfo {author} {\bibfnamefont {K.~N.}\ \bibnamefont
  {{Abazajian}}}, \bibinfo {author} {\bibfnamefont {S.}~\bibnamefont
  {{Blanchet}}}, \ and\ \bibinfo {author} {\bibfnamefont {J.~P.}\ \bibnamefont
  {{Harding}}},\ }\href {\doibase 10.1103/PhysRevD.84.103007} {\bibfield
  {journal} {\bibinfo  {journal} {\prd}\ }\textbf {\bibinfo {volume} {84}},\
  \bibinfo {eid} {103007} (\bibinfo {year} {2011})},\ \Eprint
  {http://arxiv.org/abs/1012.1247} {arXiv:1012.1247} \BibitemShut {NoStop}%
\bibitem [{\citenamefont {{Ajello}}\ \emph {et~al.}(2012)\citenamefont
  {{Ajello}}, \citenamefont {{Shaw}}, \citenamefont {{Romani}}, \citenamefont
  {{Dermer}}, \citenamefont {{Costamante}}, \citenamefont {{King}},
  \citenamefont {{Max-Moerbeck}}, \citenamefont {{Readhead}}, \citenamefont
  {{Reimer}}, \citenamefont {{Richards}},\ and\ \citenamefont
  {{Stevenson}}}]{2012ApJ...751..108A}%
  \BibitemOpen
  \bibfield  {author} {\bibinfo {author} {\bibfnamefont {M.}~\bibnamefont
  {{Ajello}}}, \bibinfo {author} {\bibfnamefont {M.~S.}\ \bibnamefont
  {{Shaw}}}, \bibinfo {author} {\bibfnamefont {R.~W.}\ \bibnamefont
  {{Romani}}}, \bibinfo {author} {\bibfnamefont {C.~D.}\ \bibnamefont
  {{Dermer}}}, \bibinfo {author} {\bibfnamefont {L.}~\bibnamefont
  {{Costamante}}}, \bibinfo {author} {\bibfnamefont {O.~G.}\ \bibnamefont
  {{King}}}, \bibinfo {author} {\bibfnamefont {W.}~\bibnamefont
  {{Max-Moerbeck}}}, \bibinfo {author} {\bibfnamefont {A.}~\bibnamefont
  {{Readhead}}}, \bibinfo {author} {\bibfnamefont {A.}~\bibnamefont
  {{Reimer}}}, \bibinfo {author} {\bibfnamefont {J.~L.}\ \bibnamefont
  {{Richards}}}, \ and\ \bibinfo {author} {\bibfnamefont {M.}~\bibnamefont
  {{Stevenson}}},\ }\href {\doibase 10.1088/0004-637X/751/2/108} {\bibfield
  {journal} {\bibinfo  {journal} {\apj}\ }\textbf {\bibinfo {volume} {751}},\
  \bibinfo {eid} {108} (\bibinfo {year} {2012})},\ \Eprint
  {http://arxiv.org/abs/1110.3787} {arXiv:1110.3787} \BibitemShut {NoStop}%
\bibitem [{\citenamefont {{Singal}}\ \emph {et~al.}(2012)\citenamefont
  {{Singal}}, \citenamefont {{Petrosian}},\ and\ \citenamefont
  {{Ajello}}}]{2012ApJ...753...45S}%
  \BibitemOpen
  \bibfield  {author} {\bibinfo {author} {\bibfnamefont {J.}~\bibnamefont
  {{Singal}}}, \bibinfo {author} {\bibfnamefont {V.}~\bibnamefont
  {{Petrosian}}}, \ and\ \bibinfo {author} {\bibfnamefont {M.}~\bibnamefont
  {{Ajello}}},\ }\href {\doibase 10.1088/0004-637X/753/1/45} {\bibfield
  {journal} {\bibinfo  {journal} {\apj}\ }\textbf {\bibinfo {volume} {753}},\
  \bibinfo {eid} {45} (\bibinfo {year} {2012})},\ \Eprint
  {http://arxiv.org/abs/1106.3111} {arXiv:1106.3111} \BibitemShut {NoStop}%
\bibitem [{\citenamefont {{Acero}}\ \emph {et~al.}(2015)\citenamefont
  {{Acero}}, \citenamefont {{Ackermann}}, \citenamefont {{Ajello}},
  \citenamefont {{Albert}}, \citenamefont {{Atwood}} \emph
  {et~al.}}]{2015ApJS..218...23A}%
  \BibitemOpen
  \bibfield  {author} {\bibinfo {author} {\bibfnamefont {F.}~\bibnamefont
  {{Acero}}}, \bibinfo {author} {\bibfnamefont {M.}~\bibnamefont
  {{Ackermann}}}, \bibinfo {author} {\bibfnamefont {M.}~\bibnamefont
  {{Ajello}}}, \bibinfo {author} {\bibfnamefont {A.}~\bibnamefont {{Albert}}},
  \bibinfo {author} {\bibfnamefont {W.~B.}\ \bibnamefont {{Atwood}}},  \emph
  {et~al.},\ }\href {\doibase 10.1088/0067-0049/218/2/23} {\bibfield  {journal}
  {\bibinfo  {journal} {\apjs}\ }\textbf {\bibinfo {volume} {218}},\ \bibinfo
  {eid} {23} (\bibinfo {year} {2015})}\BibitemShut {NoStop}%
\bibitem [{\citenamefont {Massaro}\ \emph {et~al.}(2015)\citenamefont
  {Massaro}, \citenamefont {Thompson},\ and\ \citenamefont
  {Ferrara}}]{Massaro2015}%
  \BibitemOpen
  \bibfield  {author} {\bibinfo {author} {\bibfnamefont {F.}~\bibnamefont
  {Massaro}}, \bibinfo {author} {\bibfnamefont {D.~J.}\ \bibnamefont
  {Thompson}}, \ and\ \bibinfo {author} {\bibfnamefont {E.~C.}\ \bibnamefont
  {Ferrara}},\ }\href {\doibase 10.1007/s00159-015-0090-6} {\bibfield
  {journal} {\bibinfo  {journal} {The Astronomy and Astrophysics Review}\
  }\textbf {\bibinfo {volume} {24}},\ \bibinfo {pages} {2} (\bibinfo {year}
  {2015})}\BibitemShut {NoStop}%
\bibitem [{\citenamefont {{Inoue}}(2011)}]{2011ApJ...733...66I}%
  \BibitemOpen
  \bibfield  {author} {\bibinfo {author} {\bibfnamefont {Y.}~\bibnamefont
  {{Inoue}}},\ }\href {\doibase 10.1088/0004-637X/733/1/66} {\bibfield
  {journal} {\bibinfo  {journal} {\apj}\ }\textbf {\bibinfo {volume} {733}},\
  \bibinfo {eid} {66} (\bibinfo {year} {2011})},\ \Eprint
  {http://arxiv.org/abs/1103.3946} {arXiv:1103.3946 [astro-ph.HE]} \BibitemShut
  {NoStop}%
\bibitem [{\citenamefont {{Di Mauro}}\ \emph
  {et~al.}(2014{\natexlab{a}})\citenamefont {{Di Mauro}}, \citenamefont
  {{Donato}}, \citenamefont {{Lamanna}}, \citenamefont {{Sanchez}},\ and\
  \citenamefont {{Serpico}}}]{2014ApJ...786..129D}%
  \BibitemOpen
  \bibfield  {author} {\bibinfo {author} {\bibfnamefont {M.}~\bibnamefont {{Di
  Mauro}}}, \bibinfo {author} {\bibfnamefont {F.}~\bibnamefont {{Donato}}},
  \bibinfo {author} {\bibfnamefont {G.}~\bibnamefont {{Lamanna}}}, \bibinfo
  {author} {\bibfnamefont {D.~A.}\ \bibnamefont {{Sanchez}}}, \ and\ \bibinfo
  {author} {\bibfnamefont {P.~D.}\ \bibnamefont {{Serpico}}},\ }\href {\doibase
  10.1088/0004-637X/786/2/129} {\bibfield  {journal} {\bibinfo  {journal}
  {\apj}\ }\textbf {\bibinfo {volume} {786}},\ \bibinfo {eid} {129} (\bibinfo
  {year} {2014}{\natexlab{a}})},\ \Eprint {http://arxiv.org/abs/1311.5708}
  {arXiv:1311.5708 [astro-ph.HE]} \BibitemShut {NoStop}%
\bibitem [{\citenamefont {{Di Mauro}}\ \emph
  {et~al.}(2014{\natexlab{b}})\citenamefont {{Di Mauro}}, \citenamefont
  {{Calore}}, \citenamefont {{Donato}}, \citenamefont {{Ajello}},\ and\
  \citenamefont {{Latronico}}}]{2014ApJ...780..161D}%
  \BibitemOpen
  \bibfield  {author} {\bibinfo {author} {\bibfnamefont {M.}~\bibnamefont {{Di
  Mauro}}}, \bibinfo {author} {\bibfnamefont {F.}~\bibnamefont {{Calore}}},
  \bibinfo {author} {\bibfnamefont {F.}~\bibnamefont {{Donato}}}, \bibinfo
  {author} {\bibfnamefont {M.}~\bibnamefont {{Ajello}}}, \ and\ \bibinfo
  {author} {\bibfnamefont {L.}~\bibnamefont {{Latronico}}},\ }\href {\doibase
  10.1088/0004-637X/780/2/161} {\bibfield  {journal} {\bibinfo  {journal}
  {\apj}\ }\textbf {\bibinfo {volume} {780}},\ \bibinfo {eid} {161} (\bibinfo
  {year} {2014}{\natexlab{b}})},\ \Eprint {http://arxiv.org/abs/1304.0908}
  {arXiv:1304.0908 [astro-ph.HE]} \BibitemShut {NoStop}%
\bibitem [{\citenamefont {{Calore}}\ \emph {et~al.}(2014)\citenamefont
  {{Calore}}, \citenamefont {{Di Mauro}},\ and\ \citenamefont
  {{Donato}}}]{2014ApJ...796...14C}%
  \BibitemOpen
  \bibfield  {author} {\bibinfo {author} {\bibfnamefont {F.}~\bibnamefont
  {{Calore}}}, \bibinfo {author} {\bibfnamefont {M.}~\bibnamefont {{Di
  Mauro}}}, \ and\ \bibinfo {author} {\bibfnamefont {F.}~\bibnamefont
  {{Donato}}},\ }\href {\doibase 10.1088/0004-637X/796/1/14} {\bibfield
  {journal} {\bibinfo  {journal} {\apj}\ }\textbf {\bibinfo {volume} {796}},\
  \bibinfo {eid} {14} (\bibinfo {year} {2014})},\ \Eprint
  {http://arxiv.org/abs/1406.2706} {arXiv:1406.2706 [astro-ph.HE]} \BibitemShut
  {NoStop}%
\bibitem [{\citenamefont {{Tamborra}}\ \emph {et~al.}(2014)\citenamefont
  {{Tamborra}}, \citenamefont {{Ando}},\ and\ \citenamefont
  {{Murase}}}]{2014JCAP...09..043T}%
  \BibitemOpen
  \bibfield  {author} {\bibinfo {author} {\bibfnamefont {I.}~\bibnamefont
  {{Tamborra}}}, \bibinfo {author} {\bibfnamefont {S.}~\bibnamefont {{Ando}}},
  \ and\ \bibinfo {author} {\bibfnamefont {K.}~\bibnamefont {{Murase}}},\
  }\href {\doibase 10.1088/1475-7516/2014/09/043} {\bibfield  {journal}
  {\bibinfo  {journal} {\jcap}\ }\textbf {\bibinfo {volume} {9}},\ \bibinfo
  {eid} {043} (\bibinfo {year} {2014})},\ \Eprint
  {http://arxiv.org/abs/1404.1189} {arXiv:1404.1189 [astro-ph.HE]} \BibitemShut
  {NoStop}%
\bibitem [{\citenamefont {{Ajello}}\ \emph {et~al.}(2015)\citenamefont
  {{Ajello}}, \citenamefont {{Gasparrini}}, \citenamefont
  {{S{\'a}nchez-Conde}}, \citenamefont {{Zaharijas}}, \citenamefont
  {{Gustafsson}} \emph {et~al.}}]{2015ApJ...800L..27A}%
  \BibitemOpen
  \bibfield  {author} {\bibinfo {author} {\bibfnamefont {M.}~\bibnamefont
  {{Ajello}}}, \bibinfo {author} {\bibfnamefont {D.}~\bibnamefont
  {{Gasparrini}}}, \bibinfo {author} {\bibfnamefont {M.}~\bibnamefont
  {{S{\'a}nchez-Conde}}}, \bibinfo {author} {\bibfnamefont {G.}~\bibnamefont
  {{Zaharijas}}}, \bibinfo {author} {\bibfnamefont {M.}~\bibnamefont
  {{Gustafsson}}},  \emph {et~al.},\ }\href {\doibase
  10.1088/2041-8205/800/2/L27} {\bibfield  {journal} {\bibinfo  {journal}
  {\apjl}\ }\textbf {\bibinfo {volume} {800}},\ \bibinfo {eid} {L27} (\bibinfo
  {year} {2015})},\ \Eprint {http://arxiv.org/abs/1501.05301} {arXiv:1501.05301
  [astro-ph.HE]} \BibitemShut {NoStop}%
\bibitem [{\citenamefont {Ullio}\ \emph {et~al.}(2002)\citenamefont {Ullio},
  \citenamefont {Bergstrom}, \citenamefont {Edsjo},\ and\ \citenamefont
  {Lacey}}]{Ullio:2002pj}%
  \BibitemOpen
  \bibfield  {author} {\bibinfo {author} {\bibfnamefont {P.}~\bibnamefont
  {Ullio}}, \bibinfo {author} {\bibfnamefont {L.}~\bibnamefont {Bergstrom}},
  \bibinfo {author} {\bibfnamefont {J.}~\bibnamefont {Edsjo}}, \ and\ \bibinfo
  {author} {\bibfnamefont {C.~G.}\ \bibnamefont {Lacey}},\ }\href {\doibase
  10.1103/PhysRevD.66.123502} {\bibfield  {journal} {\bibinfo  {journal} {Phys.
  Rev.}\ }\textbf {\bibinfo {volume} {D66}},\ \bibinfo {pages} {123502}
  (\bibinfo {year} {2002})},\ \Eprint {http://arxiv.org/abs/astro-ph/0207125}
  {arXiv:astro-ph/0207125 [astro-ph]} \BibitemShut {NoStop}%
\bibitem [{\citenamefont {Abazajian}\ \emph {et~al.}(2012)\citenamefont
  {Abazajian}, \citenamefont {Blanchet},\ and\ \citenamefont
  {Harding}}]{Abazajian:2010zb}%
  \BibitemOpen
  \bibfield  {author} {\bibinfo {author} {\bibfnamefont {K.~N.}\ \bibnamefont
  {Abazajian}}, \bibinfo {author} {\bibfnamefont {S.}~\bibnamefont {Blanchet}},
  \ and\ \bibinfo {author} {\bibfnamefont {J.~P.}\ \bibnamefont {Harding}},\
  }\href {\doibase 10.1103/PhysRevD.85.043509} {\bibfield  {journal} {\bibinfo
  {journal} {Phys. Rev.}\ }\textbf {\bibinfo {volume} {D85}},\ \bibinfo {pages}
  {043509} (\bibinfo {year} {2012})},\ \Eprint {http://arxiv.org/abs/1011.5090}
  {arXiv:1011.5090 [hep-ph]} \BibitemShut {NoStop}%
\bibitem [{\citenamefont {Papucci}\ and\ \citenamefont
  {Strumia}(2010)}]{Papucci:2009gd}%
  \BibitemOpen
  \bibfield  {author} {\bibinfo {author} {\bibfnamefont {M.}~\bibnamefont
  {Papucci}}\ and\ \bibinfo {author} {\bibfnamefont {A.}~\bibnamefont
  {Strumia}},\ }\href {\doibase 10.1088/1475-7516/2010/03/014} {\bibfield
  {journal} {\bibinfo  {journal} {JCAP}\ }\textbf {\bibinfo {volume} {1003}},\
  \bibinfo {pages} {014} (\bibinfo {year} {2010})},\ \Eprint
  {http://arxiv.org/abs/0912.0742} {arXiv:0912.0742 [hep-ph]} \BibitemShut
  {NoStop}%
\bibitem [{\citenamefont {{Cirelli}}\ \emph {et~al.}(2010)\citenamefont
  {{Cirelli}}, \citenamefont {{Panci}},\ and\ \citenamefont
  {{Serpico}}}]{2010NuPhB.840..284C}%
  \BibitemOpen
  \bibfield  {author} {\bibinfo {author} {\bibfnamefont {M.}~\bibnamefont
  {{Cirelli}}}, \bibinfo {author} {\bibfnamefont {P.}~\bibnamefont {{Panci}}},
  \ and\ \bibinfo {author} {\bibfnamefont {P.~D.}\ \bibnamefont {{Serpico}}},\
  }\href {\doibase 10.1016/j.nuclphysb.2010.07.010} {\bibfield  {journal}
  {\bibinfo  {journal} {Nuclear Physics B}\ }\textbf {\bibinfo {volume}
  {840}},\ \bibinfo {pages} {284} (\bibinfo {year} {2010})},\ \Eprint
  {http://arxiv.org/abs/0912.0663} {arXiv:0912.0663} \BibitemShut {NoStop}%
\bibitem [{\citenamefont {Bringmann}\ \emph {et~al.}(2014)\citenamefont
  {Bringmann}, \citenamefont {Calore}, \citenamefont {Di~Mauro},\ and\
  \citenamefont {Donato}}]{Calore:2013yia}%
  \BibitemOpen
  \bibfield  {author} {\bibinfo {author} {\bibfnamefont {T.}~\bibnamefont
  {Bringmann}}, \bibinfo {author} {\bibfnamefont {F.}~\bibnamefont {Calore}},
  \bibinfo {author} {\bibfnamefont {M.}~\bibnamefont {Di~Mauro}}, \ and\
  \bibinfo {author} {\bibfnamefont {F.}~\bibnamefont {Donato}},\ }\href
  {\doibase 10.1103/PhysRevD.89.023012} {\bibfield  {journal} {\bibinfo
  {journal} {Phys. Rev.}\ }\textbf {\bibinfo {volume} {D89}},\ \bibinfo {pages}
  {023012} (\bibinfo {year} {2014})},\ \Eprint {http://arxiv.org/abs/1303.3284}
  {arXiv:1303.3284 [astro-ph.CO]} \BibitemShut {NoStop}%
\bibitem [{\citenamefont {Di~Mauro}\ and\ \citenamefont
  {Donato}(2015)}]{DiMauro:2015tfa}%
  \BibitemOpen
  \bibfield  {author} {\bibinfo {author} {\bibfnamefont {M.}~\bibnamefont
  {Di~Mauro}}\ and\ \bibinfo {author} {\bibfnamefont {F.}~\bibnamefont
  {Donato}},\ }\href {\doibase 10.1103/PhysRevD.91.123001} {\bibfield
  {journal} {\bibinfo  {journal} {Phys. Rev.}\ }\textbf {\bibinfo {volume}
  {D91}},\ \bibinfo {pages} {123001} (\bibinfo {year} {2015})},\ \Eprint
  {http://arxiv.org/abs/1501.05316} {arXiv:1501.05316 [astro-ph.HE]}
  \BibitemShut {NoStop}%
\bibitem [{\citenamefont {Ackermann}\ \emph {et~al.}(2015)\citenamefont
  {Ackermann} \emph {et~al.}}]{Ackermann:2015tah}%
  \BibitemOpen
  \bibfield  {author} {\bibinfo {author} {\bibfnamefont {M.}~\bibnamefont
  {Ackermann}} \emph {et~al.} (\bibinfo {collaboration} {Fermi-LAT}),\ }\href
  {\doibase 10.1088/1475-7516/2015/09/008} {\bibfield  {journal} {\bibinfo
  {journal} {JCAP}\ }\textbf {\bibinfo {volume} {1509}},\ \bibinfo {pages}
  {008} (\bibinfo {year} {2015})},\ \Eprint {http://arxiv.org/abs/1501.05464}
  {arXiv:1501.05464 [astro-ph.CO]} \BibitemShut {NoStop}%
\bibitem [{\citenamefont {{Dodelson}}\ \emph {et~al.}(2009)\citenamefont
  {{Dodelson}}, \citenamefont {{Belikov}}, \citenamefont {{Hooper}},\ and\
  \citenamefont {{Serpico}}}]{2009PhRvD..80h3504D}%
  \BibitemOpen
  \bibfield  {author} {\bibinfo {author} {\bibfnamefont {S.}~\bibnamefont
  {{Dodelson}}}, \bibinfo {author} {\bibfnamefont {A.~V.}\ \bibnamefont
  {{Belikov}}}, \bibinfo {author} {\bibfnamefont {D.}~\bibnamefont {{Hooper}}},
  \ and\ \bibinfo {author} {\bibfnamefont {P.}~\bibnamefont {{Serpico}}},\
  }\href {\doibase 10.1103/PhysRevD.80.083504} {\bibfield  {journal} {\bibinfo
  {journal} {\prd}\ }\textbf {\bibinfo {volume} {80}},\ \bibinfo {eid} {083504}
  (\bibinfo {year} {2009})},\ \Eprint {http://arxiv.org/abs/0903.2829}
  {arXiv:0903.2829 [astro-ph.CO]} \BibitemShut {NoStop}%
\bibitem [{\citenamefont {{Lee}}\ \emph {et~al.}(2009)\citenamefont {{Lee}},
  \citenamefont {{Ando}},\ and\ \citenamefont
  {{Kamionkowski}}}]{2009JCAP...07..007L}%
  \BibitemOpen
  \bibfield  {author} {\bibinfo {author} {\bibfnamefont {S.~K.}\ \bibnamefont
  {{Lee}}}, \bibinfo {author} {\bibfnamefont {S.}~\bibnamefont {{Ando}}}, \
  and\ \bibinfo {author} {\bibfnamefont {M.}~\bibnamefont {{Kamionkowski}}},\
  }\href {\doibase 10.1088/1475-7516/2009/07/007} {\bibfield  {journal}
  {\bibinfo  {journal} {\jcap}\ }\textbf {\bibinfo {volume} {7}},\ \bibinfo
  {eid} {007} (\bibinfo {year} {2009})},\ \Eprint
  {http://arxiv.org/abs/0810.1284} {arXiv:0810.1284} \BibitemShut {NoStop}%
\bibitem [{\citenamefont {{Malyshev}}\ and\ \citenamefont
  {{Hogg}}(2011)}]{2011ApJ...738..181M}%
  \BibitemOpen
  \bibfield  {author} {\bibinfo {author} {\bibfnamefont {D.}~\bibnamefont
  {{Malyshev}}}\ and\ \bibinfo {author} {\bibfnamefont {D.~W.}\ \bibnamefont
  {{Hogg}}},\ }\href {\doibase 10.1088/0004-637X/738/2/181} {\bibfield
  {journal} {\bibinfo  {journal} {\apj}\ }\textbf {\bibinfo {volume} {738}},\
  \bibinfo {eid} {181} (\bibinfo {year} {2011})}\BibitemShut {NoStop}%
\bibitem [{\citenamefont {{Feyereisen}}\ \emph {et~al.}(2015)\citenamefont
  {{Feyereisen}}, \citenamefont {{Ando}},\ and\ \citenamefont
  {{Lee}}}]{2015JCAP...09..027F}%
  \BibitemOpen
  \bibfield  {author} {\bibinfo {author} {\bibfnamefont {M.~R.}\ \bibnamefont
  {{Feyereisen}}}, \bibinfo {author} {\bibfnamefont {S.}~\bibnamefont
  {{Ando}}}, \ and\ \bibinfo {author} {\bibfnamefont {S.~K.}\ \bibnamefont
  {{Lee}}},\ }\href {\doibase 10.1088/1475-7516/2015/09/027} {\bibfield
  {journal} {\bibinfo  {journal} {\jcap}\ }\textbf {\bibinfo {volume} {9}},\
  \bibinfo {eid} {027} (\bibinfo {year} {2015})},\ \Eprint
  {http://arxiv.org/abs/1506.05118} {arXiv:1506.05118} \BibitemShut {NoStop}%
\bibitem [{\citenamefont {{Massari}}\ \emph {et~al.}(2015)\citenamefont
  {{Massari}}, \citenamefont {{Izaguirre}}, \citenamefont {{Essig}},
  \citenamefont {{Albert}}, \citenamefont {{Bloom}},\ and\ \citenamefont
  {{G{\'o}mez-Vargas}}}]{2015PhRvD..91h3539M}%
  \BibitemOpen
  \bibfield  {author} {\bibinfo {author} {\bibfnamefont {A.}~\bibnamefont
  {{Massari}}}, \bibinfo {author} {\bibfnamefont {E.}~\bibnamefont
  {{Izaguirre}}}, \bibinfo {author} {\bibfnamefont {R.}~\bibnamefont
  {{Essig}}}, \bibinfo {author} {\bibfnamefont {A.}~\bibnamefont {{Albert}}},
  \bibinfo {author} {\bibfnamefont {E.}~\bibnamefont {{Bloom}}}, \ and\
  \bibinfo {author} {\bibfnamefont {G.~A.}\ \bibnamefont
  {{G{\'o}mez-Vargas}}},\ }\href {\doibase 10.1103/PhysRevD.91.083539}
  {\bibfield  {journal} {\bibinfo  {journal} {\prd}\ }\textbf {\bibinfo
  {volume} {91}},\ \bibinfo {eid} {083539} (\bibinfo {year} {2015})},\ \Eprint
  {http://arxiv.org/abs/1503.07169} {arXiv:1503.07169 [hep-ph]} \BibitemShut
  {NoStop}%
\bibitem [{\citenamefont {{Selig}}\ \emph {et~al.}(2015)\citenamefont
  {{Selig}}, \citenamefont {{Vacca}}, \citenamefont {{Oppermann}},\ and\
  \citenamefont {{En{\ss}lin}}}]{2015A&A...581A.126S}%
  \BibitemOpen
  \bibfield  {author} {\bibinfo {author} {\bibfnamefont {M.}~\bibnamefont
  {{Selig}}}, \bibinfo {author} {\bibfnamefont {V.}~\bibnamefont {{Vacca}}},
  \bibinfo {author} {\bibfnamefont {N.}~\bibnamefont {{Oppermann}}}, \ and\
  \bibinfo {author} {\bibfnamefont {T.~A.}\ \bibnamefont {{En{\ss}lin}}},\
  }\href {\doibase 10.1051/0004-6361/201425172} {\bibfield  {journal} {\bibinfo
   {journal} {\aap}\ }\textbf {\bibinfo {volume} {581}},\ \bibinfo {eid} {A126}
  (\bibinfo {year} {2015})}\BibitemShut {NoStop}%
\bibitem [{\citenamefont {{Lee}}\ \emph {et~al.}(2016)\citenamefont {{Lee}},
  \citenamefont {{Lisanti}}, \citenamefont {{Safdi}}, \citenamefont
  {{Slatyer}},\ and\ \citenamefont {{Xue}}}]{2016PhRvL.116e1103L}%
  \BibitemOpen
  \bibfield  {author} {\bibinfo {author} {\bibfnamefont {S.~K.}\ \bibnamefont
  {{Lee}}}, \bibinfo {author} {\bibfnamefont {M.}~\bibnamefont {{Lisanti}}},
  \bibinfo {author} {\bibfnamefont {B.~R.}\ \bibnamefont {{Safdi}}}, \bibinfo
  {author} {\bibfnamefont {T.~R.}\ \bibnamefont {{Slatyer}}}, \ and\ \bibinfo
  {author} {\bibfnamefont {W.}~\bibnamefont {{Xue}}},\ }\href {\doibase
  10.1103/PhysRevLett.116.051103} {\bibfield  {journal} {\bibinfo  {journal}
  {Physical Review Letters}\ }\textbf {\bibinfo {volume} {116}},\ \bibinfo
  {eid} {051103} (\bibinfo {year} {2016})}\BibitemShut {NoStop}%
\bibitem [{\citenamefont {{Lisanti}}\ \emph {et~al.}(2016)\citenamefont
  {{Lisanti}}, \citenamefont {{Mishra-Sharma}}, \citenamefont {{Necib}},\ and\
  \citenamefont {{Safdi}}}]{2016ApJ...832..117L}%
  \BibitemOpen
  \bibfield  {author} {\bibinfo {author} {\bibfnamefont {M.}~\bibnamefont
  {{Lisanti}}}, \bibinfo {author} {\bibfnamefont {S.}~\bibnamefont
  {{Mishra-Sharma}}}, \bibinfo {author} {\bibfnamefont {L.}~\bibnamefont
  {{Necib}}}, \ and\ \bibinfo {author} {\bibfnamefont {B.~R.}\ \bibnamefont
  {{Safdi}}},\ }\href {\doibase 10.3847/0004-637X/832/2/117} {\bibfield
  {journal} {\bibinfo  {journal} {\apj}\ }\textbf {\bibinfo {volume} {832}},\
  \bibinfo {eid} {117} (\bibinfo {year} {2016})}\BibitemShut {NoStop}%
\bibitem [{\citenamefont {{Zechlin}}\ \emph
  {et~al.}(2016{\natexlab{a}})\citenamefont {{Zechlin}}, \citenamefont
  {{Cuoco}}, \citenamefont {{Donato}}, \citenamefont {{Fornengo}},\ and\
  \citenamefont {{Vittino}}}]{2016ApJS..225...18Z}%
  \BibitemOpen
  \bibfield  {author} {\bibinfo {author} {\bibfnamefont {H.-S.}\ \bibnamefont
  {{Zechlin}}}, \bibinfo {author} {\bibfnamefont {A.}~\bibnamefont {{Cuoco}}},
  \bibinfo {author} {\bibfnamefont {F.}~\bibnamefont {{Donato}}}, \bibinfo
  {author} {\bibfnamefont {N.}~\bibnamefont {{Fornengo}}}, \ and\ \bibinfo
  {author} {\bibfnamefont {A.}~\bibnamefont {{Vittino}}},\ }\href {\doibase
  10.3847/0067-0049/225/2/18} {\bibfield  {journal} {\bibinfo  {journal}
  {\apjs}\ }\textbf {\bibinfo {volume} {225}},\ \bibinfo {eid} {18} (\bibinfo
  {year} {2016}{\natexlab{a}})}\BibitemShut {NoStop}%
\bibitem [{\citenamefont {{Zechlin}}\ \emph
  {et~al.}(2016{\natexlab{b}})\citenamefont {{Zechlin}}, \citenamefont
  {{Cuoco}}, \citenamefont {{Donato}}, \citenamefont {{Fornengo}},\ and\
  \citenamefont {{Regis}}}]{2016ApJ...826L..31Z}%
  \BibitemOpen
  \bibfield  {author} {\bibinfo {author} {\bibfnamefont {H.-S.}\ \bibnamefont
  {{Zechlin}}}, \bibinfo {author} {\bibfnamefont {A.}~\bibnamefont {{Cuoco}}},
  \bibinfo {author} {\bibfnamefont {F.}~\bibnamefont {{Donato}}}, \bibinfo
  {author} {\bibfnamefont {N.}~\bibnamefont {{Fornengo}}}, \ and\ \bibinfo
  {author} {\bibfnamefont {M.}~\bibnamefont {{Regis}}},\ }\href {\doibase
  10.3847/2041-8205/826/2/L31} {\bibfield  {journal} {\bibinfo  {journal}
  {\apjl}\ }\textbf {\bibinfo {volume} {826}},\ \bibinfo {eid} {L31} (\bibinfo
  {year} {2016}{\natexlab{b}})}\BibitemShut {NoStop}%
\bibitem [{\citenamefont {Wolleben}(2007)}]{Wolleben:2007pq}%
  \BibitemOpen
  \bibfield  {author} {\bibinfo {author} {\bibfnamefont {M.}~\bibnamefont
  {Wolleben}},\ }\href {\doibase 10.1086/518711} {\bibfield  {journal}
  {\bibinfo  {journal} {Astrophys. J.}\ }\textbf {\bibinfo {volume} {664}},\
  \bibinfo {pages} {349} (\bibinfo {year} {2007})},\ \Eprint
  {http://arxiv.org/abs/0704.0276} {arXiv:0704.0276 [astro-ph]} \BibitemShut
  {NoStop}%
\bibitem [{\citenamefont {Ackermann}\ \emph {et~al.}(2014)\citenamefont
  {Ackermann} \emph {et~al.}}]{Fermi-LAT:2014sfa}%
  \BibitemOpen
  \bibfield  {author} {\bibinfo {author} {\bibfnamefont {M.}~\bibnamefont
  {Ackermann}} \emph {et~al.} (\bibinfo {collaboration} {Fermi-LAT}),\ }\href
  {\doibase 10.1088/0004-637X/793/1/64} {\bibfield  {journal} {\bibinfo
  {journal} {Astrophys. J.}\ }\textbf {\bibinfo {volume} {793}},\ \bibinfo
  {pages} {64} (\bibinfo {year} {2014})},\ \Eprint
  {http://arxiv.org/abs/1407.7905} {arXiv:1407.7905 [astro-ph.HE]} \BibitemShut
  {NoStop}%
\bibitem [{\citenamefont {{Ackermann}}\ \emph
  {et~al.}(2012{\natexlab{b}})\citenamefont {{Ackermann}}, \citenamefont
  {{Ajello}}, \citenamefont {{Atwood}}, \citenamefont {{Baldini}},
  \citenamefont {{Barbiellini}} \emph {et~al.}}]{2012ApJ...761...91A}%
  \BibitemOpen
  \bibfield  {author} {\bibinfo {author} {\bibfnamefont {M.}~\bibnamefont
  {{Ackermann}}}, \bibinfo {author} {\bibfnamefont {M.}~\bibnamefont
  {{Ajello}}}, \bibinfo {author} {\bibfnamefont {W.~B.}\ \bibnamefont
  {{Atwood}}}, \bibinfo {author} {\bibfnamefont {L.}~\bibnamefont {{Baldini}}},
  \bibinfo {author} {\bibfnamefont {G.}~\bibnamefont {{Barbiellini}}},  \emph
  {et~al.},\ }\href {\doibase 10.1088/0004-637X/761/2/91} {\bibfield  {journal}
  {\bibinfo  {journal} {\apj}\ }\textbf {\bibinfo {volume} {761}},\ \bibinfo
  {eid} {91} (\bibinfo {year} {2012}{\natexlab{b}})},\ \Eprint
  {http://arxiv.org/abs/1205.6474} {arXiv:1205.6474 [astro-ph.CO]} \BibitemShut
  {NoStop}%
\bibitem [{\citenamefont {Charles}\ \emph {et~al.}(2016)\citenamefont {Charles}
  \emph {et~al.}}]{Charles:2016pgz}%
  \BibitemOpen
  \bibfield  {author} {\bibinfo {author} {\bibfnamefont {E.}~\bibnamefont
  {Charles}} \emph {et~al.} (\bibinfo {collaboration} {Fermi-LAT}),\ }\href
  {\doibase 10.1016/j.physrep.2016.05.001} {\bibfield  {journal} {\bibinfo
  {journal} {Phys. Rept.}\ }\textbf {\bibinfo {volume} {636}},\ \bibinfo
  {pages} {1} (\bibinfo {year} {2016})},\ \Eprint
  {http://arxiv.org/abs/1605.02016} {arXiv:1605.02016 [astro-ph.HE]}
  \BibitemShut {NoStop}%
\bibitem [{\citenamefont {{Acero}}\ \emph {et~al.}(2016)\citenamefont
  {{Acero}}, \citenamefont {{Ackermann}}, \citenamefont {{Ajello}},
  \citenamefont {{Albert}}, \citenamefont {{Baldini}} \emph
  {et~al.}}]{2016ApJS..223...26A}%
  \BibitemOpen
  \bibfield  {author} {\bibinfo {author} {\bibfnamefont {F.}~\bibnamefont
  {{Acero}}}, \bibinfo {author} {\bibfnamefont {M.}~\bibnamefont
  {{Ackermann}}}, \bibinfo {author} {\bibfnamefont {M.}~\bibnamefont
  {{Ajello}}}, \bibinfo {author} {\bibfnamefont {A.}~\bibnamefont {{Albert}}},
  \bibinfo {author} {\bibfnamefont {L.}~\bibnamefont {{Baldini}}},  \emph
  {et~al.},\ }\href {\doibase 10.3847/0067-0049/223/2/26} {\bibfield  {journal}
  {\bibinfo  {journal} {\apjs}\ }\textbf {\bibinfo {volume} {223}},\ \bibinfo
  {eid} {26} (\bibinfo {year} {2016})},\ \Eprint
  {http://arxiv.org/abs/1602.07246} {arXiv:1602.07246 [astro-ph.HE]}
  \BibitemShut {NoStop}%
\bibitem [{\citenamefont {{Ackermann}}\ \emph
  {et~al.}(2012{\natexlab{c}})\citenamefont {{Ackermann}}, \citenamefont
  {{Ajello}}, \citenamefont {{Atwood}}, \citenamefont {{Baldini}},
  \citenamefont {{Ballet}} \emph {et~al.}}]{2012ApJ...750....3A}%
  \BibitemOpen
  \bibfield  {author} {\bibinfo {author} {\bibfnamefont {M.}~\bibnamefont
  {{Ackermann}}}, \bibinfo {author} {\bibfnamefont {M.}~\bibnamefont
  {{Ajello}}}, \bibinfo {author} {\bibfnamefont {W.~B.}\ \bibnamefont
  {{Atwood}}}, \bibinfo {author} {\bibfnamefont {L.}~\bibnamefont {{Baldini}}},
  \bibinfo {author} {\bibfnamefont {J.}~\bibnamefont {{Ballet}}},  \emph
  {et~al.},\ }\href {\doibase 10.1088/0004-637X/750/1/3} {\bibfield  {journal}
  {\bibinfo  {journal} {\apj}\ }\textbf {\bibinfo {volume} {750}},\ \bibinfo
  {eid} {3} (\bibinfo {year} {2012}{\natexlab{c}})},\ \Eprint
  {http://arxiv.org/abs/1202.4039} {arXiv:1202.4039 [astro-ph.HE]} \BibitemShut
  {NoStop}%
\bibitem [{\citenamefont {Catena}\ and\ \citenamefont
  {Ullio}(2010)}]{Catena:2009mf}%
  \BibitemOpen
  \bibfield  {author} {\bibinfo {author} {\bibfnamefont {R.}~\bibnamefont
  {Catena}}\ and\ \bibinfo {author} {\bibfnamefont {P.}~\bibnamefont {Ullio}},\
  }\href {\doibase 10.1088/1475-7516/2010/08/004} {\bibfield  {journal}
  {\bibinfo  {journal} {JCAP}\ }\textbf {\bibinfo {volume} {1008}},\ \bibinfo
  {pages} {004} (\bibinfo {year} {2010})},\ \Eprint
  {http://arxiv.org/abs/0907.0018} {arXiv:0907.0018 [astro-ph.CO]} \BibitemShut
  {NoStop}%
\bibitem [{\citenamefont {Read}(2014)}]{Read:2014qva}%
  \BibitemOpen
  \bibfield  {author} {\bibinfo {author} {\bibfnamefont {J.~I.}\ \bibnamefont
  {Read}},\ }\href {\doibase 10.1088/0954-3899/41/6/063101} {\bibfield
  {journal} {\bibinfo  {journal} {J. Phys.}\ }\textbf {\bibinfo {volume}
  {G41}},\ \bibinfo {pages} {063101} (\bibinfo {year} {2014})},\ \Eprint
  {http://arxiv.org/abs/1404.1938} {arXiv:1404.1938 [astro-ph.GA]} \BibitemShut
  {NoStop}%
\bibitem [{\citenamefont {{Zechlin}}\ \emph {et~al.}(2012)\citenamefont
  {{Zechlin}}, \citenamefont {{Fernandes}}, \citenamefont {{Els{\"a}sser}},\
  and\ \citenamefont {{Horns}}}]{2012A&A...538A..93Z}%
  \BibitemOpen
  \bibfield  {author} {\bibinfo {author} {\bibfnamefont {H.-S.}\ \bibnamefont
  {{Zechlin}}}, \bibinfo {author} {\bibfnamefont {M.~V.}\ \bibnamefont
  {{Fernandes}}}, \bibinfo {author} {\bibfnamefont {D.}~\bibnamefont
  {{Els{\"a}sser}}}, \ and\ \bibinfo {author} {\bibfnamefont {D.}~\bibnamefont
  {{Horns}}},\ }\href {\doibase 10.1051/0004-6361/201117655} {\bibfield
  {journal} {\bibinfo  {journal} {\aap}\ }\textbf {\bibinfo {volume} {538}},\
  \bibinfo {eid} {A93} (\bibinfo {year} {2012})},\ \Eprint
  {http://arxiv.org/abs/1111.3514} {arXiv:1111.3514 [astro-ph.HE]} \BibitemShut
  {NoStop}%
\bibitem [{\citenamefont {{Ackermann}}\ \emph
  {et~al.}(2012{\natexlab{d}})\citenamefont {{Ackermann}}, \citenamefont
  {{Albert}}, \citenamefont {{Baldini}}, \citenamefont {{Ballet}},
  \citenamefont {{Barbiellini}} \emph {et~al.}}]{2012ApJ...747..121A}%
  \BibitemOpen
  \bibfield  {author} {\bibinfo {author} {\bibfnamefont {M.}~\bibnamefont
  {{Ackermann}}}, \bibinfo {author} {\bibfnamefont {A.}~\bibnamefont
  {{Albert}}}, \bibinfo {author} {\bibfnamefont {L.}~\bibnamefont {{Baldini}}},
  \bibinfo {author} {\bibfnamefont {J.}~\bibnamefont {{Ballet}}}, \bibinfo
  {author} {\bibfnamefont {G.}~\bibnamefont {{Barbiellini}}},  \emph {et~al.},\
  }\href {\doibase 10.1088/0004-637X/747/2/121} {\bibfield  {journal} {\bibinfo
   {journal} {\apj}\ }\textbf {\bibinfo {volume} {747}},\ \bibinfo {eid} {121}
  (\bibinfo {year} {2012}{\natexlab{d}})},\ \Eprint
  {http://arxiv.org/abs/1201.2691} {arXiv:1201.2691 [astro-ph.HE]} \BibitemShut
  {NoStop}%
\bibitem [{\citenamefont {{Zechlin}}\ and\ \citenamefont
  {{Horns}}(2012)}]{2012JCAP...11..050Z}%
  \BibitemOpen
  \bibfield  {author} {\bibinfo {author} {\bibfnamefont {H.-S.}\ \bibnamefont
  {{Zechlin}}}\ and\ \bibinfo {author} {\bibfnamefont {D.}~\bibnamefont
  {{Horns}}},\ }\href {\doibase 10.1088/1475-7516/2012/11/050} {\bibfield
  {journal} {\bibinfo  {journal} {\jcap}\ }\textbf {\bibinfo {volume} {11}},\
  \bibinfo {eid} {050} (\bibinfo {year} {2012})},\ \Eprint
  {http://arxiv.org/abs/1210.3852} {arXiv:1210.3852 [astro-ph.HE]} \BibitemShut
  {NoStop}%
\bibitem [{\citenamefont {{Bertoni}}\ \emph {et~al.}(2015)\citenamefont
  {{Bertoni}}, \citenamefont {{Hooper}},\ and\ \citenamefont
  {{Linden}}}]{2015JCAP...12..035B}%
  \BibitemOpen
  \bibfield  {author} {\bibinfo {author} {\bibfnamefont {B.}~\bibnamefont
  {{Bertoni}}}, \bibinfo {author} {\bibfnamefont {D.}~\bibnamefont {{Hooper}}},
  \ and\ \bibinfo {author} {\bibfnamefont {T.}~\bibnamefont {{Linden}}},\
  }\href {\doibase 10.1088/1475-7516/2015/12/035} {\bibfield  {journal}
  {\bibinfo  {journal} {\jcap}\ }\textbf {\bibinfo {volume} {12}},\ \bibinfo
  {eid} {035} (\bibinfo {year} {2015})},\ \Eprint
  {http://arxiv.org/abs/1504.02087} {arXiv:1504.02087 [astro-ph.HE]}
  \BibitemShut {NoStop}%
\bibitem [{\citenamefont {{Schoonenberg}}\ \emph {et~al.}(2016)\citenamefont
  {{Schoonenberg}}, \citenamefont {{Gaskins}}, \citenamefont {{Bertone}},\ and\
  \citenamefont {{Diemand}}}]{2016JCAP...05..028S}%
  \BibitemOpen
  \bibfield  {author} {\bibinfo {author} {\bibfnamefont {D.}~\bibnamefont
  {{Schoonenberg}}}, \bibinfo {author} {\bibfnamefont {J.}~\bibnamefont
  {{Gaskins}}}, \bibinfo {author} {\bibfnamefont {G.}~\bibnamefont
  {{Bertone}}}, \ and\ \bibinfo {author} {\bibfnamefont {J.}~\bibnamefont
  {{Diemand}}},\ }\href {\doibase 10.1088/1475-7516/2016/05/028} {\bibfield
  {journal} {\bibinfo  {journal} {\jcap}\ }\textbf {\bibinfo {volume} {5}},\
  \bibinfo {eid} {028} (\bibinfo {year} {2016})},\ \Eprint
  {http://arxiv.org/abs/1601.06781} {arXiv:1601.06781 [astro-ph.HE]}
  \BibitemShut {NoStop}%
\bibitem [{\citenamefont {{Hooper}}\ and\ \citenamefont
  {{Witte}}(2017)}]{2017JCAP...04..018H}%
  \BibitemOpen
  \bibfield  {author} {\bibinfo {author} {\bibfnamefont {D.}~\bibnamefont
  {{Hooper}}}\ and\ \bibinfo {author} {\bibfnamefont {S.~J.}\ \bibnamefont
  {{Witte}}},\ }\href {\doibase 10.1088/1475-7516/2017/04/018} {\bibfield
  {journal} {\bibinfo  {journal} {\jcap}\ }\textbf {\bibinfo {volume} {4}},\
  \bibinfo {eid} {018} (\bibinfo {year} {2017})},\ \Eprint
  {http://arxiv.org/abs/1610.07587} {arXiv:1610.07587 [astro-ph.HE]}
  \BibitemShut {NoStop}%
\bibitem [{\citenamefont {Calore}\ \emph {et~al.}(2017)\citenamefont {Calore},
  \citenamefont {De~Romeri}, \citenamefont {Di~Mauro}, \citenamefont {Donato},\
  and\ \citenamefont {Marinacci}}]{Calore:2016ogv}%
  \BibitemOpen
  \bibfield  {author} {\bibinfo {author} {\bibfnamefont {F.}~\bibnamefont
  {Calore}}, \bibinfo {author} {\bibfnamefont {V.}~\bibnamefont {De~Romeri}},
  \bibinfo {author} {\bibfnamefont {M.}~\bibnamefont {Di~Mauro}}, \bibinfo
  {author} {\bibfnamefont {F.}~\bibnamefont {Donato}}, \ and\ \bibinfo {author}
  {\bibfnamefont {F.}~\bibnamefont {Marinacci}},\ }\href {\doibase
  10.1103/PhysRevD.96.063009} {\bibfield  {journal} {\bibinfo  {journal} {Phys.
  Rev.}\ }\textbf {\bibinfo {volume} {D96}},\ \bibinfo {pages} {063009}
  (\bibinfo {year} {2017})},\ \Eprint {http://arxiv.org/abs/1611.03503}
  {arXiv:1611.03503 [astro-ph.HE]} \BibitemShut {NoStop}%
\bibitem [{\citenamefont {{Einasto}}(1965)}]{1965TrAlm...5...87E}%
  \BibitemOpen
  \bibfield  {author} {\bibinfo {author} {\bibfnamefont {J.}~\bibnamefont
  {{Einasto}}},\ }\href@noop {} {\bibfield  {journal} {\bibinfo  {journal}
  {Trudy Astrofizicheskogo Instituta Alma-Ata}\ }\textbf {\bibinfo {volume}
  {5}},\ \bibinfo {pages} {87} (\bibinfo {year} {1965})}\BibitemShut {NoStop}%
\bibitem [{\citenamefont {Bringmann}\ and\ \citenamefont
  {Weniger}(2012)}]{Bringmann:2012ez}%
  \BibitemOpen
  \bibfield  {author} {\bibinfo {author} {\bibfnamefont {T.}~\bibnamefont
  {Bringmann}}\ and\ \bibinfo {author} {\bibfnamefont {C.}~\bibnamefont
  {Weniger}},\ }\href {\doibase 10.1016/j.dark.2012.10.005} {\bibfield
  {journal} {\bibinfo  {journal} {Phys. Dark Univ.}\ }\textbf {\bibinfo
  {volume} {1}},\ \bibinfo {pages} {194} (\bibinfo {year} {2012})},\ \Eprint
  {http://arxiv.org/abs/1208.5481} {arXiv:1208.5481 [hep-ph]} \BibitemShut
  {NoStop}%
\bibitem [{\citenamefont {{Cirelli}}\ \emph {et~al.}(2011)\citenamefont
  {{Cirelli}}, \citenamefont {{Corcella}}, \citenamefont {{Hektor}},
  \citenamefont {{H{\"u}tsi}}, \citenamefont {{Kadastik}}, \citenamefont
  {{Panci}}, \citenamefont {{Raidal}}, \citenamefont {{Sala}},\ and\
  \citenamefont {{Strumia}}}]{2011JCAP...03..051C}%
  \BibitemOpen
  \bibfield  {author} {\bibinfo {author} {\bibfnamefont {M.}~\bibnamefont
  {{Cirelli}}}, \bibinfo {author} {\bibfnamefont {G.}~\bibnamefont
  {{Corcella}}}, \bibinfo {author} {\bibfnamefont {A.}~\bibnamefont
  {{Hektor}}}, \bibinfo {author} {\bibfnamefont {G.}~\bibnamefont
  {{H{\"u}tsi}}}, \bibinfo {author} {\bibfnamefont {M.}~\bibnamefont
  {{Kadastik}}}, \bibinfo {author} {\bibfnamefont {P.}~\bibnamefont {{Panci}}},
  \bibinfo {author} {\bibfnamefont {M.}~\bibnamefont {{Raidal}}}, \bibinfo
  {author} {\bibfnamefont {F.}~\bibnamefont {{Sala}}}, \ and\ \bibinfo {author}
  {\bibfnamefont {A.}~\bibnamefont {{Strumia}}},\ }\href {\doibase
  10.1088/1475-7516/2011/03/051} {\bibfield  {journal} {\bibinfo  {journal}
  {\jcap}\ }\textbf {\bibinfo {volume} {3}},\ \bibinfo {eid} {051} (\bibinfo
  {year} {2011})},\ \Eprint {http://arxiv.org/abs/1012.4515} {arXiv:1012.4515
  [hep-ph]} \BibitemShut {NoStop}%
\bibitem [{Note1()}]{Note1}%
  \BibitemOpen
  \bibinfo {note} {\protect \emph {Fermi}-LAT data are publicly available at
  \protect \texttt
  {https://heasarc.gsfc.nasa.gov/FTP/fermi/data/lat/\\weekly/photon/}}\BibitemShut
  {NoStop}%
\bibitem [{Note2()}]{Note2}%
  \BibitemOpen
  \bibinfo {note} {\protect \texttt
  {https://fermi.gsfc.nasa.gov/ssc/data/analysis/\\software/}}\BibitemShut
  {NoStop}%
\bibitem [{\citenamefont {{G{\'o}rski}}\ \emph {et~al.}(2005)\citenamefont
  {{G{\'o}rski}}, \citenamefont {{Hivon}}, \citenamefont {{Banday}},
  \citenamefont {{Wandelt}}, \citenamefont {{Hansen}}, \citenamefont
  {{Reinecke}},\ and\ \citenamefont {{Bartelmann}}}]{2005ApJ...622..759G}%
  \BibitemOpen
  \bibfield  {author} {\bibinfo {author} {\bibfnamefont {K.~M.}\ \bibnamefont
  {{G{\'o}rski}}}, \bibinfo {author} {\bibfnamefont {E.}~\bibnamefont
  {{Hivon}}}, \bibinfo {author} {\bibfnamefont {A.~J.}\ \bibnamefont
  {{Banday}}}, \bibinfo {author} {\bibfnamefont {B.~D.}\ \bibnamefont
  {{Wandelt}}}, \bibinfo {author} {\bibfnamefont {F.~K.}\ \bibnamefont
  {{Hansen}}}, \bibinfo {author} {\bibfnamefont {M.}~\bibnamefont
  {{Reinecke}}}, \ and\ \bibinfo {author} {\bibfnamefont {M.}~\bibnamefont
  {{Bartelmann}}},\ }\href {\doibase 10.1086/427976} {\bibfield  {journal}
  {\bibinfo  {journal} {\apj}\ }\textbf {\bibinfo {volume} {622}},\ \bibinfo
  {pages} {759} (\bibinfo {year} {2005})},\ \Eprint
  {http://arxiv.org/abs/astro-ph/0409513} {astro-ph/0409513} \BibitemShut
  {NoStop}%
\bibitem [{\citenamefont {{Feroz}}\ and\ \citenamefont
  {{Hobson}}(2008)}]{2008MNRAS.384..449F}%
  \BibitemOpen
  \bibfield  {author} {\bibinfo {author} {\bibfnamefont {F.}~\bibnamefont
  {{Feroz}}}\ and\ \bibinfo {author} {\bibfnamefont {M.~P.}\ \bibnamefont
  {{Hobson}}},\ }\href {\doibase 10.1111/j.1365-2966.2007.12353.x} {\bibfield
  {journal} {\bibinfo  {journal} {\mnras}\ }\textbf {\bibinfo {volume} {384}},\
  \bibinfo {pages} {449} (\bibinfo {year} {2008})},\ \Eprint
  {http://arxiv.org/abs/0704.3704} {arXiv:0704.3704} \BibitemShut {NoStop}%
\bibitem [{\citenamefont {{Feroz}}\ \emph {et~al.}(2009)\citenamefont
  {{Feroz}}, \citenamefont {{Hobson}},\ and\ \citenamefont
  {{Bridges}}}]{2009MNRAS.398.1601F}%
  \BibitemOpen
  \bibfield  {author} {\bibinfo {author} {\bibfnamefont {F.}~\bibnamefont
  {{Feroz}}}, \bibinfo {author} {\bibfnamefont {M.~P.}\ \bibnamefont
  {{Hobson}}}, \ and\ \bibinfo {author} {\bibfnamefont {M.}~\bibnamefont
  {{Bridges}}},\ }\href {\doibase 10.1111/j.1365-2966.2009.14548.x} {\bibfield
  {journal} {\bibinfo  {journal} {\mnras}\ }\textbf {\bibinfo {volume} {398}},\
  \bibinfo {pages} {1601} (\bibinfo {year} {2009})},\ \Eprint
  {http://arxiv.org/abs/0809.3437} {arXiv:0809.3437} \BibitemShut {NoStop}%
\bibitem [{\citenamefont {{Rolke}}\ \emph {et~al.}(2005)\citenamefont
  {{Rolke}}, \citenamefont {{L{\'o}pez}},\ and\ \citenamefont
  {{Conrad}}}]{2005NIMPA.551..493R}%
  \BibitemOpen
  \bibfield  {author} {\bibinfo {author} {\bibfnamefont {W.~A.}\ \bibnamefont
  {{Rolke}}}, \bibinfo {author} {\bibfnamefont {A.~M.}\ \bibnamefont
  {{L{\'o}pez}}}, \ and\ \bibinfo {author} {\bibfnamefont {J.}~\bibnamefont
  {{Conrad}}},\ }\href {\doibase 10.1016/j.nima.2005.05.068} {\bibfield
  {journal} {\bibinfo  {journal} {Nuclear Instruments and Methods in Physics
  Research A}\ }\textbf {\bibinfo {volume} {551}},\ \bibinfo {pages} {493}
  (\bibinfo {year} {2005})},\ \Eprint {http://arxiv.org/abs/physics/0403059}
  {physics/0403059} \BibitemShut {NoStop}%
\bibitem [{\citenamefont {{Su}}\ \emph {et~al.}(2010)\citenamefont {{Su}},
  \citenamefont {{Slatyer}},\ and\ \citenamefont
  {{Finkbeiner}}}]{2010ApJ...724.1044S}%
  \BibitemOpen
  \bibfield  {author} {\bibinfo {author} {\bibfnamefont {M.}~\bibnamefont
  {{Su}}}, \bibinfo {author} {\bibfnamefont {T.~R.}\ \bibnamefont {{Slatyer}}},
  \ and\ \bibinfo {author} {\bibfnamefont {D.~P.}\ \bibnamefont
  {{Finkbeiner}}},\ }\href {\doibase 10.1088/0004-637X/724/2/1044} {\bibfield
  {journal} {\bibinfo  {journal} {\apj}\ }\textbf {\bibinfo {volume} {724}},\
  \bibinfo {pages} {1044} (\bibinfo {year} {2010})},\ \Eprint
  {http://arxiv.org/abs/1005.5480} {arXiv:1005.5480 [astro-ph.HE]} \BibitemShut
  {NoStop}%
\bibitem [{\citenamefont {{Ackermann}}\ \emph {et~al.}(2014)\citenamefont
  {{Ackermann}}, \citenamefont {{Albert}}, \citenamefont {{Atwood}},
  \citenamefont {{Baldini}}, \citenamefont {{Ballet}} \emph
  {et~al.}}]{2014ApJ...793...64A}%
  \BibitemOpen
  \bibfield  {author} {\bibinfo {author} {\bibfnamefont {M.}~\bibnamefont
  {{Ackermann}}}, \bibinfo {author} {\bibfnamefont {A.}~\bibnamefont
  {{Albert}}}, \bibinfo {author} {\bibfnamefont {W.~B.}\ \bibnamefont
  {{Atwood}}}, \bibinfo {author} {\bibfnamefont {L.}~\bibnamefont {{Baldini}}},
  \bibinfo {author} {\bibfnamefont {J.}~\bibnamefont {{Ballet}}},  \emph
  {et~al.},\ }\href {\doibase 10.1088/0004-637X/793/1/64} {\bibfield  {journal}
  {\bibinfo  {journal} {\apj}\ }\textbf {\bibinfo {volume} {793}},\ \bibinfo
  {eid} {64} (\bibinfo {year} {2014})},\ \Eprint
  {http://arxiv.org/abs/1407.7905} {arXiv:1407.7905 [astro-ph.HE]} \BibitemShut
  {NoStop}%
\bibitem [{\citenamefont {{Casandjian}}\ \emph {et~al.}(2009)\citenamefont
  {{Casandjian}}, \citenamefont {{Grenier}},\ and\ \citenamefont {{for the
  Fermi Large Area Telescope Collaboration}}}]{2009arXiv0912.3478C}%
  \BibitemOpen
  \bibfield  {author} {\bibinfo {author} {\bibfnamefont {J.-M.}\ \bibnamefont
  {{Casandjian}}}, \bibinfo {author} {\bibfnamefont {I.}~\bibnamefont
  {{Grenier}}}, \ and\ \bibinfo {author} {\bibnamefont {{for the Fermi Large
  Area Telescope Collaboration}}},\ }\href@noop {} {\bibfield  {journal}
  {\bibinfo  {journal} {ArXiv e-prints}\ } (\bibinfo {year} {2009})},\ \Eprint
  {http://arxiv.org/abs/0912.3478} {arXiv:0912.3478 [astro-ph.HE]} \BibitemShut
  {NoStop}%
\bibitem [{\citenamefont {Daylan}\ \emph {et~al.}(2016)\citenamefont {Daylan},
  \citenamefont {Finkbeiner}, \citenamefont {Hooper}, \citenamefont {Linden},
  \citenamefont {Portillo}, \citenamefont {Rodd},\ and\ \citenamefont
  {Slatyer}}]{Daylan:2014rsa}%
  \BibitemOpen
  \bibfield  {author} {\bibinfo {author} {\bibfnamefont {T.}~\bibnamefont
  {Daylan}}, \bibinfo {author} {\bibfnamefont {D.~P.}\ \bibnamefont
  {Finkbeiner}}, \bibinfo {author} {\bibfnamefont {D.}~\bibnamefont {Hooper}},
  \bibinfo {author} {\bibfnamefont {T.}~\bibnamefont {Linden}}, \bibinfo
  {author} {\bibfnamefont {S.~K.~N.}\ \bibnamefont {Portillo}}, \bibinfo
  {author} {\bibfnamefont {N.~L.}\ \bibnamefont {Rodd}}, \ and\ \bibinfo
  {author} {\bibfnamefont {T.~R.}\ \bibnamefont {Slatyer}},\ }\href {\doibase
  10.1016/j.dark.2015.12.005} {\bibfield  {journal} {\bibinfo  {journal} {Phys.
  Dark Univ.}\ }\textbf {\bibinfo {volume} {12}},\ \bibinfo {pages} {1}
  (\bibinfo {year} {2016})},\ \Eprint {http://arxiv.org/abs/1402.6703}
  {arXiv:1402.6703 [astro-ph.HE]} \BibitemShut {NoStop}%
\bibitem [{\citenamefont {Calore}\ \emph {et~al.}(2015)\citenamefont {Calore},
  \citenamefont {Cholis},\ and\ \citenamefont {Weniger}}]{Calore:2014xka}%
  \BibitemOpen
  \bibfield  {author} {\bibinfo {author} {\bibfnamefont {F.}~\bibnamefont
  {Calore}}, \bibinfo {author} {\bibfnamefont {I.}~\bibnamefont {Cholis}}, \
  and\ \bibinfo {author} {\bibfnamefont {C.}~\bibnamefont {Weniger}},\ }\href
  {\doibase 10.1088/1475-7516/2015/03/038} {\bibfield  {journal} {\bibinfo
  {journal} {JCAP}\ }\textbf {\bibinfo {volume} {1503}},\ \bibinfo {pages}
  {038} (\bibinfo {year} {2015})},\ \Eprint {http://arxiv.org/abs/1409.0042}
  {arXiv:1409.0042 [astro-ph.CO]} \BibitemShut {NoStop}%
\bibitem [{\citenamefont {Cohen}\ \emph {et~al.}(2017)\citenamefont {Cohen},
  \citenamefont {Murase}, \citenamefont {Rodd}, \citenamefont {Safdi},\ and\
  \citenamefont {Soreq}}]{Cohen:2016uyg}%
  \BibitemOpen
  \bibfield  {author} {\bibinfo {author} {\bibfnamefont {T.}~\bibnamefont
  {Cohen}}, \bibinfo {author} {\bibfnamefont {K.}~\bibnamefont {Murase}},
  \bibinfo {author} {\bibfnamefont {N.~L.}\ \bibnamefont {Rodd}}, \bibinfo
  {author} {\bibfnamefont {B.~R.}\ \bibnamefont {Safdi}}, \ and\ \bibinfo
  {author} {\bibfnamefont {Y.}~\bibnamefont {Soreq}},\ }\href {\doibase
  10.1103/PhysRevLett.119.021102} {\bibfield  {journal} {\bibinfo  {journal}
  {Phys. Rev. Lett.}\ }\textbf {\bibinfo {volume} {119}},\ \bibinfo {pages}
  {021102} (\bibinfo {year} {2017})},\ \Eprint
  {http://arxiv.org/abs/1612.05638} {arXiv:1612.05638 [hep-ph]} \BibitemShut
  {NoStop}%
\bibitem [{\citenamefont {Linden}\ \emph {et~al.}(2016)\citenamefont {Linden},
  \citenamefont {Rodd}, \citenamefont {Safdi},\ and\ \citenamefont
  {Slatyer}}]{Linden:2016rcf}%
  \BibitemOpen
  \bibfield  {author} {\bibinfo {author} {\bibfnamefont {T.}~\bibnamefont
  {Linden}}, \bibinfo {author} {\bibfnamefont {N.~L.}\ \bibnamefont {Rodd}},
  \bibinfo {author} {\bibfnamefont {B.~R.}\ \bibnamefont {Safdi}}, \ and\
  \bibinfo {author} {\bibfnamefont {T.~R.}\ \bibnamefont {Slatyer}},\ }\href
  {\doibase 10.1103/PhysRevD.94.103013} {\bibfield  {journal} {\bibinfo
  {journal} {Phys. Rev.}\ }\textbf {\bibinfo {volume} {D94}},\ \bibinfo {pages}
  {103013} (\bibinfo {year} {2016})},\ \Eprint
  {http://arxiv.org/abs/1604.01026} {arXiv:1604.01026 [astro-ph.HE]}
  \BibitemShut {NoStop}%
\bibitem [{\citenamefont {{Ackermann}}\ \emph {et~al.}(2015)\citenamefont
  {{Ackermann}}, \citenamefont {{Albert}}, \citenamefont {{Anderson}},
  \citenamefont {{Atwood}}, \citenamefont {{Baldini}} \emph
  {et~al.}}]{2015PhRvL.115w1301A}%
  \BibitemOpen
  \bibfield  {author} {\bibinfo {author} {\bibfnamefont {M.}~\bibnamefont
  {{Ackermann}}}, \bibinfo {author} {\bibfnamefont {A.}~\bibnamefont
  {{Albert}}}, \bibinfo {author} {\bibfnamefont {B.}~\bibnamefont
  {{Anderson}}}, \bibinfo {author} {\bibfnamefont {W.~B.}\ \bibnamefont
  {{Atwood}}}, \bibinfo {author} {\bibfnamefont {L.}~\bibnamefont {{Baldini}}},
   \emph {et~al.},\ }\href {\doibase 10.1103/PhysRevLett.115.231301} {\bibfield
   {journal} {\bibinfo  {journal} {Physical Review Letters}\ }\textbf {\bibinfo
  {volume} {115}},\ \bibinfo {eid} {231301} (\bibinfo {year} {2015})},\ \Eprint
  {http://arxiv.org/abs/1503.02641} {arXiv:1503.02641 [astro-ph.HE]}
  \BibitemShut {NoStop}%
\end{thebibliography}%

\end{document}